\documentclass[times,final]{elsarticle}

\usepackage{jcomp}
\usepackage{framed,multirow}

\usepackage{lineno,hyperref}
\usepackage[utf8]{inputenc}
\usepackage[version=4]{mhchem}
\usepackage{siunitx}
\usepackage{graphicx}
\usepackage{epstopdf, epsfig}
\usepackage{lipsum}
\usepackage{caption}
\usepackage{subcaption}
\usepackage{amsmath}
\usepackage{commath}
\usepackage{wrapfig}
\usepackage{xcolor}
\usepackage{subfiles}
\usepackage{pdfpages}
\usepackage{verbatim}
\usepackage{multirow}
\usepackage{booktabs}
\usepackage{mathabx}
\usepackage{tikz}
\usepackage{bm}
\usetikzlibrary{shapes}

\journal{Journal of Computational Physics}

\biboptions{authoryear}

\graphicspath{
{./figures/}
{./figures//FR/}
{./figures//FR/burgers/}
{./figures//FR/sod/}
{./figures//FR/semiLocalE/}
{./figures//FR/shuOsher/}
{./figures//FR/shockVortex/}
{./figures//FR/mach10/}
{./figures//FR/gfunc/}
}

\newcommand{\bookmark}[1]{{\color{red} \#\#\#\#\#\# BOOKMARK \#\#\#\#\#\# }}

\newcommand{\reviewerA}[1]{{\color{black} #1}}
\newcommand{\reviewerB}[1]{{\color{black} #1}}

\DeclareRobustCommand\sampleline[1]{%
  \tikz\draw[#1] (0,0) (0,\the\dimexpr\fontdimen22\textfont2\relax)
  -- (2em,\the\dimexpr\fontdimen22\textfont2\relax);%
}

\newcommand{\be}{\begin{equation}}
\newcommand{\ee}{\end{equation}}

\newcommand{\modelFullName}{Legendre Spectral Viscosity}
\newcommand{\modelAcronym}{LSV}

\begin{document}

\begin{frontmatter}

\title{A Legendre spectral viscosity (\modelAcronym) method applied to shock capturing for high-order flux reconstruction schemes
}
\author[aff1]{Victor C. B. Sousa\corref{mycorrespondingauthor}}
\cortext[mycorrespondingauthor]{Corresponding author}
\ead{vsousa@purdue.edu}

\verso{V. Sousa and C. Scalo}

\author[aff1,aff2]{Carlo Scalo}

\address[aff1]{School of Mechanical Engineering, Purdue University, West Lafayette, IN, 47907}
\address[aff2]{School of Aeronautics and Astronautics, Purdue University, West Lafayette, IN, 47907}

\begin{abstract}

A novel approach to shock capturing for high-order flux reconstruction schemes is derived based on the mathematical formalism of the filtered governing equations. While the latter perspective is only typically used for turbulence modeling in the context of Large-Eddy Simulations (LES), the novel \modelFullName\ (\modelAcronym) sub-filter scale (SFS) closure model is capable of performing simulations in the presence of shock-discontinuities. The \modelAcronym\ method exploits the set of hierarchical basis functions formed by the Legendre polynomials to extract the information on the energy content near the resolution limit and estimate the overall magnitude of the required SFS dissipative terms, resulting in a scheme that dynamically activates only in cells where nonlinear behavior is important. Additionally, the modulation of such terms in the Legendre spectral space allows for the concentration of the dissipative action at small scales. The proposed method is tested in canonical shock-dominated flow setups in both one and two dimensions. These include the 1D Burgers' problem, a 1D shock tube, a 1D shock-entropy wave interaction, a 2D inviscid shock-vortex interaction and a 2D double Mach reflection. Results showcase a high-degree of resolution power, achieving accurate results with a small number of degrees of freedom, and robustness, being able to capture shocks associated with the Burgers' equation and the 1D shock tube within a single cell with orders 120 and higher.
\end{abstract}


\end{frontmatter}

%-------------------------------------------------------------------------------------------------------------------------------------------------------------------------%
%!TEX root = ../SousaLSV_JCP_2022.tex

\section{Introduction} \label{sec:Introduction}

In recent years, considerable effort has been put into the development of discontinuous high-order methods capable of solving a wide range of PDE's, including the Navier-Stokes equations. Such methods display a high convergence rate and, in general, require smaller grid resolutions to achieve accurate results when compared to the low-order counterpart. The latter, however, are more robust due to the inherent dissipation embedded in the low-order numerics. The presence of high-gradients in the computational domain, such as shock-induced discontinuities, is particularly challenging for high-order schemes and dissipation must be introduced to avoid numerical instabilities. Nonetheless, this step must be performed meticulously so that the large scales present are minimally affected. A careless introduction of a dissipation term to stabilize a shock could deteriorate a large band of the spectrum of resolvable turbulence when both phenomena happen concomitantly, for example. The current work devises a method for stabilizing nonlinear equations in the context of discontinuous high-order numerical methods by extending the Large Eddy Simulation (LES) framework, based on solving the filtered quantities and modeling of the sub-filter scale (SFS) energy flux, to capturing any discontinuity present in the flow. In that way, the introduced dissipation is close to the minimum necessary to maintain the filtered solution. A novelty is introduced in the closure model, where a projection of the discontinuous solution within each element onto the set of Legendre basis functions is used to inform the magnitude of the dissipation and to modulate its strength at different scales. It is a further development built upon the Quasi-Spectral Viscosity (QSV) method \citep{Sousa_ArXiv_2021} capable of modeling shocks and turbulence at the same time using high-order finite-difference operators. 

Prior to moving forward into the novel methodology, a background in the development of discontinuous high-order methods is given. These methods approximate the solution within each cell by using a polynomial with arbitrary order, a characteristic of finite-element methods, and construct a unique upwind flux between cell interfaces, a trait of finite-volume methods. From this framework, two approaches were initially derived: the discontinuous Galerkin (DG) method \citep{Cockburn_MC_1989,bassi1997BR1high} and the staggered-grid approach \citep{kopriva1996conservative}. The first projects the solution vector onto test functions and applies the summation by parts rule to get to a weak form of the conservation equations where the boundary terms are substituted by the numerical upwind flux. The second solves the differential form of the conservation laws by using Lagrange polynomials and its properties to perform derivations and interpolations. It is named staggered-grid approach because it stores solution values at the Gauss quadrature points and flux values at the Gauss-Lobatto quadrature points. The latter include the points at the cell interface and its values are used to get the unique numerical flux between two cells. 

Despite the differences between the aforementioned methods, the flux reconstruction (FR) method \citep{huynh2007flux}, also referred to as the correction procedure via reconstruction (CPR) method \citep{Wang_JCP_2009} when applied to triangular mesh elements, encompasses both into a single framework. It does that through defining a generic correction function, $g$, that has unitary value at one interface and zero at the others. The correction function is then scaled by the difference between the discontinuous flux value interpolated from the interior points and the common numerical upwind flux at each of the interfaces in order to construct a continuous flux. The divergence of the continuous flux is used to advance the solution in time. Because the FR method is more general, the current work will focus on how to carryout sub-filter scale modeling for this framework, in the context of shock-dominated flows.

As previously mentioned, high-order methods suffer from numerical instabilities in the presence of high gradients, such as shocks. To simulate those types of flows mainly two types of strategies were developed: either use a nonlinear limiter to control the solution or add an artificial dissipation term, both active only in the vicinity of a discontinuity. The first strategy involves restricting the order of the interpolating polynomial in problematic regions and leading to the development of total variation bounded (TVB) type slope limiters \citep{Shu_MoC_1987,Wang_2002_JCP} and moment limiters \citep{Biswas_ANM_1994,Burbeau_JCP_2001,Krivodonova_JCP_2007}. Such limiters have been successful in some applications but they can induce over dampening of the solution in smooth regions \citep{Zhang_JCP_2017}. In the same category are the weighted essentially non-oscillatory (WENO) type schemes \citep{Qiu_SIAM_2005,Zhu_JCP_2008,Luo_JCP_2007}, which constrain the solution to being non-oscillatory near discontinuities with high-order stencils. These however, can also induce over dampening of physical fluctuations near discontinuities \citep{HagaKawai_JCP_2019}. 

The other strategy is related to the addition of artificial viscosity proportional to Laplacian terms in the governing equations to induce energy dissipation. \citet{persson2006sub} introduced a sensor that activated a constant artificial viscosity term inside an element which contained a discontinuity and were able to capture subcell shocks with length scale resolution in the order of $O(h/p)$, which was shown effective for obtaining RANS solutions using high-order DG approximations \citep{Nguyen_AIAA_2007}. Moving forward, \citet{Barter_JCP_2010} pointed out the drawbacks of a piecewise-constant artificial viscosity approach and proposed a smooth variation of the artificial viscosity within each element, which rendered better predictions of surface heating when solving a hypersonic flow over a sphere with a DG approach. Recently, the Local Artificial Diffusivity (LAD) scheme, also based on artificial viscosity and initially developed for high order finite difference methods \citep{Cook_PoF_2007,Kawai_JCP_2008,Kawai_JCP_2010} was extended to the spectral difference staggered-grid method \citep{Premasuthan_2014_I} and to the FR approach \citep{HagaKawai_JCP_2019}. Most notably, the FR-LAD method was able to outperform FR-WENO schemes in a shock-entropy wave interaction problem with respect to resolving the trailing small scale waves in the density field. Moreover, \citet{HagaKawai_JCP_2019} also showed a LES of an overexpanded supersonic jet displaying the capability of the method for engineering applications. Despite these results, the FR-LAD approach showed some drawbacks: first, it is limited to polynomial orders of degree 4; second, its implementation requires a smoothing filter performed through a computationally intensive restriction-prolongation operation.

In parallel, the Spectral Vanishing Viscosity (SVV) method was developed for single elements with spectral discretization based on Fourier \citep{Tadmor_SIAM_1989,Tadmor_NASA_1990} and Legendre \citep{Maday_JNA_1993} basis functions with the pretense of recovering spectral convergence properties when solving conservation laws that form shock discontinuities spontaneously. This method consists of adding a convolution kernel to the laplacian term, i.e. a wavenumber-dependent viscosity term, to control the dissipation added by concentrating it at the small scales near the resolution limit with the objective to prevent oscillations and lead to convergence to the unique entropy solution. Initially, it was shown by mathematical proofs and numerical experiments that a sharp viscosity activation after a certain wavenumber was successful in achieving the aforementioned stability goals \citep{Tadmor_SIAM_1989} but further study by  \citet{Maday_JNA_1993} showed an increased performance if a smooth convolution kernel was used. 

Subsequent work by \citet{Karamanos_JCP_2000} extended the methodology to spectral/$hp$ methods based on a DG formulation and also applied it to incompressible turbulent flows by adding the artificial wavenumber dependent dissipation term to the momentum equation without further consideration. In a following publication, \citet{Kirby_JFE_2002} applied the same framework to compressible turbulent simulations and dissipation terms were added to the mass, momentum and energy equations. Although their results show that the method works, additional analysis on the reasons why it worked were lacking. These questions were answered by \citet{Sousa_ArXiv_2021}, who showed that a low-band filtering of a nonlinear conservation law leads to a wavenumber dependent sub-filter flux term, therefore, the artificial addition of such dissipation terms on previous publications were akin to performing simulations of the filtered Navier-Stokes equations but without exploiting some of the physical aspects of this consideration. Moreover, \citet{Sousa_ArXiv_2021} presents an extension to shock discontinuities of the applicability of previous findings by \citet{kraichnan1976eddy} and \citet{chollet1981parameterization}, who analyzed turbulence spectra and found that a wavenumber-dependent eddy viscosity that smoothly transitioned from a plateau at large scales to a peak around the smallest resolvable scale was needed to model the energy transfer across a filter cutoff. Ultimately, the filtered equations interpretation also helps to explain the reason why a smooth convolution kernel led to better performance, it being closer than a sharp kernel to the physical energy flux to sub-filter scales (SFS) needed to sustain the filtered solution.

In an attempt to extend the SVV methodology to FR methods, \citet{Asthana_JCP_2015} first proved that the addition of adequate artificial dissipation can ensure nonlinear stability and chose to implement the needed dissipation as the result of a filtering operation in Fourier space via convolution. Being the convolution operation nonlocal, the elementwise solution was padded with information from its neighbors and it was argued that this was critical to stabilizing inter-element discontinuities. The proposed formulation was shown to be able to solve a 1D Burgers problem with polynomial order as high as 119 and to solve problems related to the 1D Euler equations up to order 8.

In the current work, it is proposed to leverage a different set of hierarchical basis functions, the Legendre polynomials, to introduce a novel SVV-based methodology for FR methods, which allows to localize even more the operations to single elements where high gradients are present. Moreover, the new formulation builds upon the QSV framework \citep{Sousa_ArXiv_2021} by using the information from the energy at the smallest resolvable scale to dynamically inform the magnitude of the dissipation needed and by modulating the dissipation in the Legendre spectral space. \reviewerA{In section \ref{sec:FRLSV}, the filtered equations, the sub-filter scale (SFS) flux terms and their flux reconstruction formulation are presented. Next, in section \ref{sec:LSV}, the details of the \modelFullName\ (\modelAcronym) closure for the modeled SFS stresses and its development based on the Legendre polynomials is presented. This section gathers all the information needed to implement the proposed approach.
}  %
Following, section \ref{sec:PerfLSV} starts by addressing 1D shock-related problems such as the Burgers' equation, the Sod shock tube and the Shu-Osher shock/entropy wave interaction via numerical experiments. Furthermore, two dimensional problems involving shocks and vorticity are addressed in this section. These results will show the performance of the currently proposed method in discontinuity capturing in high-order flux reconstruction settings. Although, the presented methodology is theoretically able to model turbulence and shocks concomitantly, in the example of \citet{Sousa_ArXiv_2021}, it was decided to leave the analysis of three-dimensional problems encompassing shocks, turbulence and their interaction for a future work.

%-------------------------------------------------------------------------------------------------------------------------------------------------------------------------%

%-------------------------------------------------------------------------------------------------------------------------------------------------------------------------%
%!TEX root = ../SousaLSV_JCP_2022.tex

\section{Flux reconstruction implementation} \label{sec:FRLSV}

In this section the filtered compressible Navier-Stokes equations will be presented together with a description of the sub-filter scale terms, all within the context of high-order Flux Reconstruction (FR) \citep{huynh2007flux} methods. \reviewerA{The proposed LSV closure, which possesses shock-capturing capabilities, is addressed later, in section \ref{sec:LSV}.}

\subsection{Governing equations}

The compressible Navier-Stokes system of equations is filtered by an operation that commutes with the derivation, 

\begin{equation} \label{eqn:filter}
\overline{f}({\bf x}) = \int f({\bf x}')\overline{G}({\bf x},{\bf x}')d{\bf x}',
\end{equation}

\noindent with an associated filter width ($\overline{\Delta}$). After the initial filtering, two routes could be chosen, the Reynolds-based or the Favre-based filtered equations. The choice of following the Reynolds method leads to all equations in the system requiring closure models and to the need of modeling the role of small-scale density variations separately. \citet{Sidharth_JFM_2018} argued that modeling the sub-filter scale (SFS) flux in the density field in an independent fashion could lead to an improvement in LES of variable-density turbulence. Moreover, this would be consistent with the addition of an artificial mass dissipation term as proposed by \citet{HagaKawai_JCP_2019}. On the other hand, the use of the Favre-based method leads to implicitly solving density related nonlinear terms by defining the Favre filter operation as,

\begin{equation} 
\check{f} = \frac{\overline{\rho f}}{\overline{\rho}}. 
\end{equation}

\noindent In the end, solving for the Favre filtered quantities $(\check{f})$ simplifies the LES equations for compressible flows. Moreover, compressible LES implementations based on the Favre-filtered equations were already successfully performed with different closure models by, for example, \citet{moin1991dynamic}, \citet{Normand_TCFD_1992}, \citet{vreman1995priori} and \citet{NagarajanLF_JCP_2003}. The current work chooses to follow the path of the latter, and derive the Favre-filtered Navier-Stokes relations introducing a pressure correction based on the sub-filter contribution to the velocity advection,

\begin{equation} \label{eqn:FavreCont}
\frac{\partial \overline \rho}{\partial t} + \frac{\partial \overline {\rho} \check{u}^j}{\partial x^j} = 0,
\end{equation}

\begin{equation} 
\frac{\partial \overline{\rho}  \check{u}^i}{\partial t} + \frac{\partial}{\partial x^j} \left( \overline \rho \check{u}^i \check{u}^j  +  \overline{p} \delta^{ij} -  \mu \check{\sigma}^{ij} +  \overline \rho \tau^{ij} \right) = 0,
\end{equation}

\begin{equation} \label{eqn:CoE}
\frac{\partial \overline E}{\partial t} + \frac{\partial}{\partial x^j} \left( (\overline E + \overline p) \check u^j - k \frac{\partial \check T}{\partial x^j}  - \mu \check \sigma^{ij} \check u^i +  \frac{1}{2} \left( \frac{\gamma \pi^j}{\gamma - 1}  + \overline \rho C_p q^{j} \right) +  \frac{1}{2} \overline \rho \zeta^{j}  \right) = \mu \epsilon,
\end{equation}

\begin{equation}  \label{eqn:FavreEnergy}
\frac{\overline p} {\gamma - 1} = \overline E -  \frac{1}{2} \overline \rho \check u^i \check u^i - \frac{1}{2} \overline \rho \tau^{ii}.
\end{equation}

The nonlinear terms that contribute to the energy flux from large to small scales are, the SFS stress tensor, 
\be \label{eqn:comptauij}
\tau^{ij} = \widecheck{u^i u^j} - \check u^i \check u^j,
\ee
\noindent the SFS heat flux,  

\be \label{eqn:compqj}
q^{j} = \widecheck{T u^j} - \check T \check u^j,
\ee

\noindent the SFS pressure-work,  

\be \label{eqn:comppij}
\pi^{j} = \overline{p u^j} - \overline p \check u^j,
\ee
 
\noindent the SFS kinetic energy advection, 
\be \label{eqn:compnuj}
\zeta^j = \widecheck{u^k u^k u^j} - \check u^k \check u^k\check u^j,
\ee

\noindent and the SFS heat dissipation 
\be \label{eqn:compeps}
\epsilon = \frac{\partial \overline{ \sigma^{ij}  u^i}}{\partial x^j} -  \frac{\partial \check \sigma^{ij} \check u^i}{\partial x^j}.
\ee

Sub-filter contributions resulting from the nonlinearities involving either the molecular viscosity or conductivity's dependency on temperature have been neglected following \citet{vreman1995priori}, who showed those are negligible in comparison against the other terms' magnitudes. \citet{NagarajanLF_JCP_2003} hypothesized $\zeta^j$ and $\epsilon$ are also much smaller than the remaining terms and showed good modeling capability through numerical experiments. Following these results, \citet{Sousa_ArXiv_2021} also neglected these terms and also able successfully to perform simulations of shocks, turbulence and their interaction.

A difference in the current form of the filtered compressible Navier-Stokes equations is the presence of a SFS pressure-work ($\pi^j$). In previous literature, an arbitrary choice was made to model the sub-filter components emanating from the nonlinear advection of the total energy and pressure as proportional to a sub-filter heat flux. Although this is an appropriate interpretation given the correct scaling, there is an advantage for including both the SFS pressure-work and heat flux into the formulation. This stems from the fact that a smooth temperature field does not imply the same property to the pressure field and vice-versa. Furthermore, since the gradient of a field acts as a sensor for high wavenumbers and, being the SFS contributions generally modeled as being proportional to the gradient of the field they are related to, using only the temperature field would render the final model essentially blind to spurious oscillations in the pressure field. Ultimately, the formulation in equation \eqref{eqn:CoE} averages the contributions to sub-filter scales that would arise if only the pressure or the temperature field were being used in the modeling process. This way, in theory, no additional dissipation is added into the system of equations but the SFS terms are now able to sense high-wavenumber build up on both the pressure and temperature fields.

\subsection{Flux reconstruction methodology}

\subsubsection{Grid transformation}

In a discontinuous numerical discretization, the computational domain ($\Omega$) is subdivided into non-overlapping elements ($\Omega_k$) such that

\be
\Omega = \bigcup_k \Omega_k.
\ee

\noindent Each individual cell is then mapped through an invertible relation from the physical cartesian coordinate system, ${\bf x} = (x^1,x^2,x^3)$, onto a standard element $\tilde \Omega := \{\xi^i | -1\leq \xi^i  \leq 1\}$ for $i = 1, 2$ and $3$. Note that this definition is only viable for hexahedral cells, which the current work chooses to focus on. Flux reconstruction implementations on triangles and prisms can be found in \citet{Wang_JCP_2009} and \citet{WilliamsJameson_JCP_2013}. 

Moving forward, if a generalized curvilinear grid transformation is used and the resulting equations are filtered in the reference space, one reaches the curvilinear compressible filtered Navier-Stokes equations. These, together with the aforementioned simplifications, can be written in the standard element coordinates, ${\bf \xi} = (\xi^1,\xi^2,\xi^3)$, as 

\begin{equation} \label{eqn:dens}
	\frac{\partial \overline{J \rho}}{\partial t}  + \frac{\partial }{\partial \xi^j}\left( \overline{J \rho} \check v^j\right)= 0,
\end{equation}

\begin{equation} \label{eqn:mom}
	\frac{\partial \overline{J \rho}\check v^i}{\partial t}  + \frac{\partial}{\partial \xi^j}\left(\overline{J \rho} \check v^i \check v^j + \overline{J p} g^{ij} - J \check \sigma^{ij} + \overline{J \rho} \tau^{ij}\right)= 
	-\Gamma^i_{qj}\left(\overline{J \rho} \check v^q \check v^j + \overline{J p} g^{qj} -  J \check \sigma^{qj} + \overline{J \rho} \tau^{qj}\right),
\end{equation}

\begin{equation}  \label{eqn:ene}
	\frac{\partial \overline{J E}}{\partial t}  + \frac{\partial}{\partial \xi^j}\left(\overline{J(E + p)}\check v^j + J \check Q^{j} - J \check \sigma^{ij}g_{ik}\check v^k + \frac{1}{2} \left( \frac{\gamma \pi^j}{\gamma - 1} + \overline {J \rho} C_p q^{j} \right)\right)= 0.
\end{equation}

\noindent Similar equations were first introduced by \citet{Jordan_JCP_1999} who developed the incompressible LES methodology in generalized curvilinear coordinates which were then expanded to the compressible version by \citet{nagarajan2007leading}. Here,  $J$ is the Jacobian of the transformation which is the determinant of the Jacobi matrix ($J_{ij} = \partial x^i/\partial \xi^j$), the velocity vector is mapped, $v^i =  \frac{\partial \xi^i}{\partial x^j} u^j $, and the Favre filtering is slightly modified, being

\be
\check f = \frac{\overline{J \rho f}}{\overline{J \rho}}.
\ee

\noindent Moreover, the tensors accounting for a general curvilinear mapping of the physical element to the euclidean standard cell are the covariant and contravariant metric tensors 

\be
g_{ij} = \frac{\partial x^i \partial x^j}{\partial \xi^k \partial \xi^k}, \quad g^{ij} = \frac{\partial \xi^k \partial \xi^k}{\partial x^i \partial x^j}, 
\ee

\noindent respectively, and the Christoffel symbol of the second kind, 

\be \label{eqn:Chris}
\Gamma^i_{qj} = \frac{\partial \xi^i}{\partial x^l}  \frac{\partial^2 x^l}{\partial \xi^q \partial \xi^j}. 
\ee

\noindent In the derivation of these equations, the metric tensors and Christoffel symbols are assumed to be varying slowly over the spatial support of the filter kernel, therefore leading to no additional sub filter flux terms.

Ultimately, following the inner product rules and the transformations between covariant and contravariant vector components in the space of the nonorthogonal spatially varying basis functions, the viscous stress tensor and the heat flux vector are described by

\begin{equation} \label{eqn:totEnergy}
\frac{\overline{J p}}{\gamma - 1}  = \overline{J E} -  \frac{1}{2}  \overline {J \rho} g_{ij} \check v^i \check v^j -  \frac{1}{2}  \overline {J \rho} g_{ij} \tau^{ij},
\end{equation}

\begin{equation}
	\check \sigma^{ij} = \mu \left(g^{jk} \frac{\partial \check v^i}{\partial \xi^k} + g^{ik} \frac{\partial \check v^j}{\partial \xi^k} - \frac{2}{3} g^{ij}\frac{\partial \check v^k}{\partial \xi^k} \right), 
\end{equation}

\begin{equation}
	\check Q_j = -k  g^{ij} \frac{\partial \check T}{\partial \xi^i}.
\end{equation}

The complete generic system of equations that is originated from a curvilinear mapping of a physical element to a standard cell is reported here for the sake of completeness. In practice some simplifying assumptions can be applied to the governing equations. If, for example, a linear mapping is assumed between the physical and standard elements, then the Christoffel symbols are all equal to zero due to the second derivative term present in its definition \eqref{eqn:Chris}. Furthermore, if the transformation is assumed to be orthogonal, then all the off-diagonal terms of the metric tensors $g_{ij}$ and $g^{ij}$ will also be trivially zero. In that case, the transformation is reduced to a simple scaling of the element. In fact, in 1D, the whole transformation can be condensed into a single factor calibrating the derivatives, as show in \citet{huynh2007flux}. 

\subsubsection{Numerical discretization of a reference element} \label{subsubsection:Numerical}

The conservation equations in the frame of reference of the standard element are then discretized within each cell by a set of solution points. This allows high-order accuracy calculations within the element, for example a $N$-th order polynomial interpolation can be achieved by using the product of $N+1$ points in each coordinate direction. Since a three-dimensional cell can be simply described by an outer product of 1D discretizations in each direction, the discussion hereafter will be focused on the simpler case.  

Now, take $u \in P_N$, the space of $N$-th order polynomials, as an approximation of an unknown quantity $U$. Let $\{h_j(\xi)\}_{j=0}^N$ be the set of Lagrange polynomials associated with the Gauss-Legendre quadrature points $\{\xi_j\}_{j=0}^N$.  The quadrature nodes, for this case, are defined as being the zeros of $L_{N+1}(\xi)$, where $L_N(\xi)$ is the \reviewerA{$N$-th} order Legendre polynomial. \citet{Shen_SpectralBook_2011} describes an efficient algorithm to compute the location of the interior nodes of the Legendre-Gauss by finding the eigenvalues of the following matrix,

\begin{equation}
A_{N+1} =
  \begin{bmatrix}
    0 & \sqrt{b_1}& ...& & 0\\
    \sqrt{b_1}& 0 & \sqrt{b_2} & &\vdots \\
   \vdots & \ddots & \ddots &\ddots &  \\
    & & \sqrt{b_{N-1}}& 0 & \sqrt{b_{N}}\\
    0& & ... & \sqrt{b_{N}} & 0
  \end{bmatrix},
\end{equation}

\noindent where,

\begin{equation}
b_i = \frac{i^2}{4i^2 -1}.
\end{equation}

\noindent This set of points are chosen since they influence the stability of the system by decreasing the importance of the Runge's phenomenon \citep{Karniadakis2013spectral} and minimizing aliasing errors in nonlinear flux reconstruction schemes \citep{Jameson_JSC_2012}. Then,

\be
u(\xi) = \sum_{j=0}^N u(\xi_j)h_j(\xi).
\ee

\noindent Clearly, differentiating $u(\xi)$ $m$ times just means evaluating the $m$-th derivative of the Lagrange polynomials, $\{h^{(m)}_j(\xi)\}_{j=0}^N$. This operation can be further simplified to be efficiently numerically implemented. \citet{Shen_SpectralBook_2011} has proved that the collocated $m$-th derivative can be computed as the $m$-th power of $D$, the matrix representation of the first-order derivative operation. The entries of $D$,

\begin{equation}
D_{ij} = 
\begin{cases}
\frac{L'_{N+1}(\xi_i)}{L'_{N+1}(\xi_j)}\frac{1}{\xi_i - \xi_j}, & \text{if } i \neq j ,\\
\frac{\xi_i}{1 - \xi_i^2}, & \text{if } i = j,\\
\end{cases}
\end{equation}

\noindent associated to the Gauss-Legendre quadrature points can be determined by the using the Legendre polynomials ($L_n$), which obey the following recurrence relations

\be
(n+1)L_{n+1}(\xi)=(2n+1)\xi L_n(\xi)- n L_{n-1}(\xi), 
\ee

\be
(1 - \xi^2)L'_{n}(\xi)=\frac{n(n+1)}{2n+1} \left(  L_{n-1}(\xi) -  L_{n+1}(\xi) \right),
\ee

\noindent for $n \ge 1$ and with $L_0 = 1$. Ultimately the collocated derivative operation is defined as

\be
u^{(m)}(\xi_i) = D_{ij}^m\ u(\xi_j),
\ee

\noindent where $D^m$ is the $m$-th power of the derivation matrix.
 
 At this point the particularities of the Flux Reconstruction (FR) scheme start. The governing equations for the standard element \eqref{eqn:dens}-\eqref{eqn:ene} are cast in the format of the divergence of a flux. The formulation of an inherently conservative scheme needs the flux vector to be identical at interfaces between neighboring cells. Since the FR scheme relies on independent discontinuous cells to work, the interpolated flux values from neighboring cells at the common interfaces are different in general. The flux discontinuity is addressed by defining a numerical flux function. In the current work the Lax-Friedrichs flux was used, which retains a positivity-preserving property when applied to the compressible Euler equations \citep{Zhang_JCP_2017}. The common value is then used to correct the initially discontinuous flux functions in each cell and the divergence is taken. 
 
 The distinction of the FR scheme relies on how such correction procedure is performed. Now consider a 1D formulation and let $f_k \in P_N$ be the approximation of the flux vector $F$ within the $k$-th cell. The interpolated value from the solution points to the cell interfaces can be evaluated  at the cells' interfaces via,
 
 \be
f_k(\pm1) = \sum_{j=0}^N f_k(\xi_j)h_j(\pm1).
\ee

\noindent These values, as well as the ones obtained from neighboring cells, $f_{k-1}(+1)$ and $f_{k+1}(-1)$  in this example, are used as inputs for the numerical flux function which outputs a unique flux value at each respective interface, 

\be
f^{un}_{k-\frac{1}{2}} = \text{Numerical Flux}\left[ f_{k-1}(+1) , f_k(-1) \right], \quad f^{un}_{k+\frac{1}{2}} = \text{Numerical Flux}\left[ f_{k}(+1) , f_{k+1}(-1) \right].
\ee

\noindent \reviewerA{Although various numerical flux strategies can be used in this step, the Lax-Friedrichs flux was adopted and its one dimensional version is presented hereafter for completeness: 

\be
f^{un}_{k-\frac{1}{2}} = \frac{f_{k-1}(+1) + f_k(-1) - a_{k-1,k} (U_{k} - U_{k-1})}{2}   , \quad f^{un}_{k+\frac{1}{2}} =  \frac{f_{k}(+1) + f_{k+1}(-1) - a_{k,k+1} (U_{k+1} - U_{k})}{2}.
\ee

\noindent  Here, $U$ is the conserved variable related to the corresponding flux value $f$ and $a_{\cdot,\cdot}$ is the maximum eigenvalue of the Jacobian matrix $|\frac{\partial f}{\partial U}|$ taken at a cell interface. This value is equal to the magnitude of the normal velocity crossing said interface augmented by the speed of sound, that is,

\be
a_{k-1,k} = \max_{i = k-1,k} \left(|v_i^1(\pm 1)| +  \sqrt{\frac{\gamma p_i(\pm 1)}{\rho_i(\pm 1)}} \right), \quad a_{k,k+1} = \max_{i = k,k+1} \left(|v_i^1(\pm 1)| +  \sqrt{\frac{\gamma p_i(\pm 1)}{\rho_i(\pm 1)}} \right),
\ee

\noindent when a perfect gas equation of state is considered. In higher dimensions the maximum operation is taken across all the points at a given cell interface and the velocity vector is dotted with the normal component of said interface. A detailed summary of a multidimensional implementation of such a flux, together with a description on its weak positivity property when applied to the Euler equations in high-order schemes, can be found in \citep{Zhang_JCP_2017}. Other possible choices for the flux function include the Godunov \citep{Godunov_1959} or HLLC flux schemes \citep{HLL_1983,Toro_SW_1994,Batten_SIAM_1997}. }

At this point, the FR scheme introduces correction functions for the left ($g_L$) and right ($g_R$) boundaries \citep{huynh2007flux}. These correction functions are defined as symmetric $(N +1)$th-order polynomials, i.e. $g_R(\xi)=-g_L(-\xi)$, that satisfy the boundary conditions $g_L(-1)=1$ and  $g_L(+1)=0$. These properties allow the construction of a continuous flux function $f^c_k$ by modifying the values at the interfaces as

 \be
f^c_k(\xi) = f_k(\xi) + \left(f^{un}_{k-\frac{1}{2}}-f_k(-1) \right)g_L(\xi) + \left(f^{un}_{k+\frac{1}{2}}-f_k(+1) \right)g_R(\xi).
\ee 

\noindent The divergence of the continuous flux at the solution points $\{\xi_i\}_{i=0}^N$ needed to advance the solution in time is then obtained

\be \label{eqn:DivContFlux}
\frac{\partial f^c_k}{\partial \xi}\Big|_{\xi = \xi_i} = D_{ij}f_k(\xi_j) + \left(f^{un}_{k-\frac{1}{2}}-f_k(-1) \right)g'_L(\xi_i) + \left(f^{un}_{k+\frac{1}{2}}-f_k(+1) \right)g'_R(\xi_i).
\ee 

\citet{huynh2007flux} suggested various polynomials to be used as correction functions. For example, if the right Radau polynomial,

\be
R_{R,N+1} = \frac{(-1)^{N+1}}{2}\left(L_{N+1} - L_N\right),
\ee

\noindent is used as $g_L$, then the scheme becomes equivalent to the nodal DG scheme for linear problems. With this in mind, \citet{huynh2007flux} denominated this polynomial as $g_{DG,N+1}$. Other suitable correction functions were also presented, such as

\be
g_{Ga,N+1} = \frac{N+1}{2N + 1} R_{R,N+1} + \frac{N}{2N + 1} R_{R,N},
\ee

\be
g_{2,N+1} = \frac{N}{2N + 1} R_{R,N+1} + \frac{N+1}{2N + 1} R_{R,N}.
\ee

\begin{figure}[ht]
\centering
\includegraphics[width=1.\linewidth]{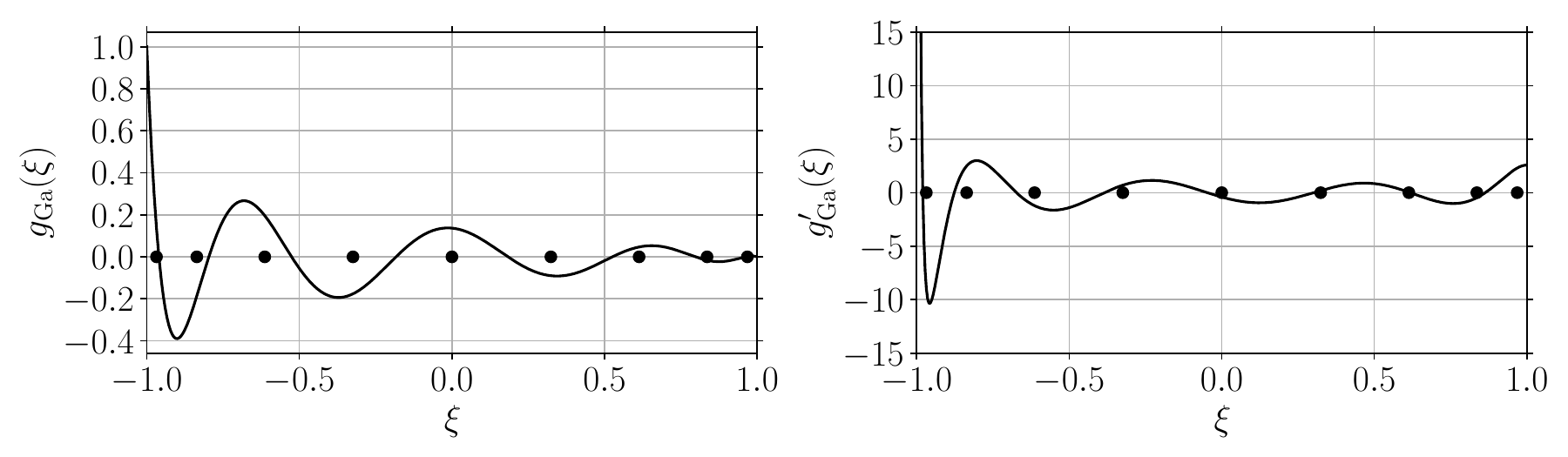}
\caption{Example of the left correction function $g_{Ga}$ and its derivative $g'_{Ga}$ when $N=8$. The Legendre-Gauss quadrature points,  $\{\xi_j\}_{j=0}^8$, are also shown.
}\label{fig:gL} 
\end{figure}

The use of different correction functions render different accuracy and stability properties to the scheme. Among these, $g_{DG}$ is the most accurate but also the one with the biggest constraint on time step size. These properties are inverted for $g_{2}$. Being a good compromise, $g_{Ga}$, shown in figure \ref{fig:gL}, is adopted in the current work. Similar motives also led \citet{HagaKawai_JCP_2019} to choose $g_{Ga}$ as a correction function. 

\reviewerA{
\section{\modelFullName\ (\modelAcronym) closure models} \label{sec:LSV}
}

The filtered compressible Navier-Stokes system of equations require a sub-filter scale (SFS) model to be closed. With only information on the large scales of the conserved variables, it is not possible to exactly derive an expression for the SFS terms $\tau^{ij}$ \eqref{eqn:comptauij}, $q^{j}$ \eqref{eqn:compqj} and $\pi^{j}$\eqref{eqn:comppij} , which need to be estimated based on their corresponding resolved field. 

Based on previous literature on spectral eddy viscosity models for LES \citep{kraichnan1976eddy,chollet1981parameterization} and spectral artificial viscosity models for shock capturing \citep{Tadmor_SIAM_1989,Tadmor_NASA_1990}, a method capable of performing both tasks simultaneously, the Quasi-Spectral Viscosity (QSV) closure, was developed by \citet{Sousa_ArXiv_2021}.  Although QSV was developed to be used in high-order finite difference methods in structure grids, the logic behind it is still useful in the current discontinuous flux reconstruction application. 

The two steps for implementing the QSV method were, first, estimating the cutoff energy, $E_N$, and second, introducing a wavenumber dependency capable of concentrating the dissipation near the filter cutoff. The same steps are performed in the current implementation with the difference that, instead of using a Fourier basis to project the solution, a Legendre basis functions is used to estimate $E_N$ and modulate the spectral viscosity. \reviewerA{The operations in the Legendre space required to perform these tasks are introduced in section \ref{subsec:LegendreOp}. The complete set of steps required for implementing the \modelAcronym~closure are discussed in section \ref{subsec:LSVimplementation}. Additionally, a deeper analysis of the spectral modulation procedure is given in section \ref{subsec:ModTransFunct}.}

Two important advantages of using the Legendre polynomials in discontinuous FR methods should be mentioned: first, the periodicity constraint related to Fourier modes is relaxed and local operations within each discontinuous cell can be performed; second, it is synergic, since the implementation of FR method is already based on Legendre polynomials, as discussed in section \ref{subsubsection:Numerical}. Additionally, \citet{Maday_JNA_1993} analyzed the Burgers' equation in the context of a Legendre pseudo-spectral method and proved that the use of a wavenumber-dependent viscosity concentrated near the highest resolvable mode led to convergence to the exact entropy solution.
\\

\subsection{Legendre Space Projection and Filtering} \label{subsec:LegendreOp}

The energy at the cutoff estimation and the modal dissipation modulation depend on the Legendre transform operation, i.e. the projection of the solution onto the hierarchical Legendre modes. This operation can be performed using the information stored on the solution points due to the exactness of the Legendre-Gauss-type quadrature, 

\begin{equation}
 \langle p,1 \rangle = \int_{-1}^1 p(\xi) d\xi =  \sum_{j=0}^N p(\xi_j) w_j, 
\end{equation}

\noindent which hold for any polynomial that belongs to the polynomial space $P$ up to order $2N+1$. Here, 

\begin{equation}
w_j = \frac{2}{(1-\xi_j^2)[L'_{N+1}(\xi_j)]^2},
\end{equation}

\noindent are the Gauss quadrature weights defined for each node $\{\xi_j\}_{j=0}^N$. This information allows the calculation of the forward discrete Legendre transform ($\mathcal{L}$), 

\be
\tilde u_n = \mathcal{L}[{\bf u}] = \frac{\langle u,L_n \rangle}{\| L_n\|^2} = \frac{1}{\gamma_n} \sum_{j=0}^N u(\xi_j) L_n(\xi_j) w_j, 
\ee

\noindent where $\gamma_n = \frac{2}{2n+1}$, through which the magnitude of the hierarchical modes ($\tilde u_n$) are found. We now define and a {\it modulated} backward discrete Legendre transform,

\be
 \mathcal{L}^{-1}[\sigma, { \bf \tilde u}]  = \sum_{n=0}^N \sigma\left(\eta\right) \tilde u_n L_n(\xi_j), 
\ee

\noindent where $\eta = n/N$, through which the modulated values in physical space can be recovered. \reviewerB{A generic modulation function, $\sigma\left(\eta\right)$ is construed such that it is nonzero only for $0 \le \eta \le 1$, therefore depending only on the resolved wavenumbers. This characteristic will allow it to be used in the current work to control the spectral behavior of the added sub filter flux models.} 

\begin{figure}[ht]
\centering
\includegraphics[width=.4\linewidth]{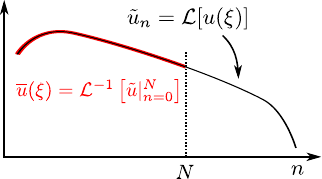}
\caption{Schematic illustration of a sharp spectral Legendre filtering operation. The horizontal axis indicates the Legendre mode number (not the Fourier-mode wavenumber).
}\label{fig:LegSharpSpectralFilter} 
\end{figure}

Additionally, note that the simple choice of restricting the Legendre modal expansion up to $N$ modes with $\sigma\left(n/N\right) = 1$  for all $n$ modes is equivalent to a sharp spectral Legendre filtering operation of a general function $u(\xi)$ to $P_N$, the space of polynomials of order up to $N$, as displayed in figure \ref{fig:LegSharpSpectralFilter}. This corresponds to the primary filter operation used in the derivation of the compressible filtered Navier-Stokes equations \eqref{eqn:dens} - \eqref{eqn:ene} when applied to the current flux reconstruction context. Moreover, if $u(\xi) \in P_N$ and $\sigma\left(n/N\right) = 1$ $\forall\ n$, then the discrete forward and backward Legendre transforms are the inverse of one another.

Although these operations seem computationally intensive, given that $L_n(\xi_j)$  $j,n=0,1,...,N$ are precomputed, the transforms can be carried out by a matrix multiplication in $N^2$ operations. Ultimately, a Legendre modulation operation is defined as a composition between a forward and a backwards transform  

\be \label{eqn:Mod}
{\mathcal{M}}[\sigma, {\bf u}] = \mathcal{L}^{-1}[\sigma, \mathcal{L}[{\bf u}]],
\ee

\noindent where $\sigma$ is a generic Legendre space modulation function, similar to the Fourier space filters defined in \citet{Gottlieb_SIAM_1997}.

\reviewerA{
\subsection{Implementation of the \modelFullName~(\modelAcronym) closure } \label{subsec:LSVimplementation} %\label{subsubsec:CutoffMod}
}

Moving ahead, the closure models for the system of equations \eqref{eqn:dens}-\eqref{eqn:ene} is addressed. First, the estimated cutoff energy for the $\overline v^j$ velocity component, $E^i_N(\overline v^j)$, can be recovered in the $i$-th direction by taking the forward Legendre transform of the kinetic energy field and storing the magnitude of the last mode, 

\be \label{eqn:spectralKE}
 E^i_{N}\left(\overline v^j \right) = \left|\frac{1}{\gamma_N} \sum_{k=0}^N \frac{1}{2}(\overline v^j(\xi^i_k))^2 L_N(\xi^i_k) w_k\right|, 
\ee

\noindent which, in turn, can be used to estimate the sub-filter velocity scale, $\upsilon^i(\overline v^j) = \sqrt{\frac{N}{2} E^i_{N}\left(\overline v^j \right)}$, where the pre-factor $N/2$ is the average grid spacing in the standard cell. This term, together with information on the local grid spacing used to estimate the sub-filter length scale ($\ell^i = \Delta^i$), can be used to inform the magnitude of the needed dissipation, 

\be \label{eqn:Dissipation}
\mathcal{D}^{ij} = \upsilon^i(\overline v^j)  \ell^j, 
\ee

\noindent in the sub-filter flux closure models. Moreover, in one dimension, the recovered information on the energy of the last mode is constant throughout a cell. A similar behavior is desirable in higher dimension formulations to mitigate spurious oscillations. With that in mind, a maximum operation is performed between the values for the energy at the cutoff obtained at different planes in a singular cell.

At this point, the background necessary to introduce the sub-filter scale (SFS) flux closure models is covered. They are written as,

\begin{eqnarray} \label{eqn:fullclosure_tau}
\tau^{lm} &=& - {\mathcal{M}}\left[\sigma_e, \frac{1}{2}\left(\mathcal{D}^{lm} \frac{\partial \overline v^l}{\partial \xi^m} + \mathcal{D}^{ml} \frac{\partial \overline v^m}{\partial \xi^l}\right)\right],\\  
 \label{eqn:fullclosure_q}
  q^{l} &=& - {\mathcal{M}}\left[ \sigma_e, \mathcal{D}^{ll} \frac{\partial \overline T}{\partial \xi^l}\right],\\
  \label{eqn:fullclosure_pi}
  \pi^{l} &=& -  \mathcal{D}^{ll} \frac{\partial \overline p}{\partial \xi^l}.
\end{eqnarray}

\begin{figure}[ht]
\centering
\includegraphics[width=1.\linewidth]{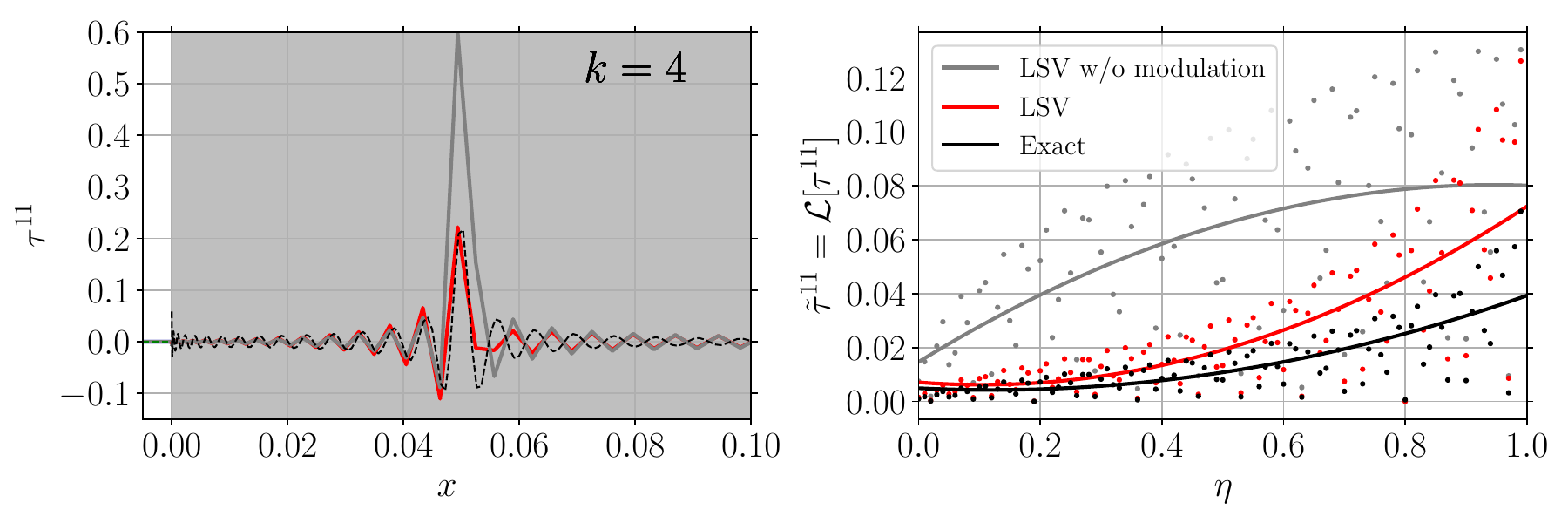}
\caption{\reviewerB{{\it A priori} analysis performed on the Burgers' equation with initial conditions given in equation \eqref{eqn:BurgersInitCond} when the periodic domain $x^1 \in [-1,1]$ is discretized with $K = 8$ cells and polynomial order of $N=100$. Only the solution at time $t = 1.05$ and in the cell $k = 4$, which contains the shock, is shown. The spectrum of the $\tau^{11}$ stress is shown in the Legendre space, $\eta$, with colored dots. A fit is also shown with a solid line to show the effects of the LSV modulation, achieving a trend closer to the exact stresses.}
}\label{fig:aprioriModulation}
\end{figure}

\noindent Here, $\sigma_e$ is the chosen modulation transfer function (see figure \ref{fig:aprioriModulation}) used to decrease the dissipation added to low wavenumbers, i.e. large scales, in the SFS stress tensor and in the SFS heat flux. The SFS pressure-work, in turn, is not modulated in the current formulation because a higher level of numerical stability, necessary when solving strong shocks, was achieved. In practice, the low wavenumber pressure dissipation helped control pressure undershoots and maintain is positivity. With the help of {\it a priori} analysis of 1D shock dominated flows the curve,

\be \label{eq:modulation_transfer_function}
\sigma_e(\eta) = 0.2 + 0.8\frac{e^{4\eta}- 1}{e^4 - 1},
\ee
 
\noindent was adopted. Although this modulation transfer function has been used successfully in numerical experiments reported in the current manuscript, it is not optimized. \reviewerA{Additional discussion on the modulation transfer function and the {\it a priori} analysis is provided hereafter in section \ref{subsec:ModTransFunct}}.

Moving forward, the last step in the implementation of the closure models in a discontinuous flux reconstruction setting is taking the derivatives related to the SFS modeling accounting for the discontinuity of the solution across neighboring cells. \reviewerA{This step is applicable to the divergence of all the modeled sub-filter scale flux terms. As an example, the divergence of the SFS heat flux, 

\be 
\frac{\partial q^l_k}{\partial \xi}\Big|_{\xi = \xi_i} = D_{ij}q^l_k(\xi_j) + \left[ q^{l,\text{avg}}_{k-\frac{1}{2}}-q^l_k(-1) \right] g'_L(\xi_i) + \left[ q^{l,\text{avg}}_{k+\frac{1}{2}}-q^l_k(+1) \right] g'_R(\xi_i),
\ee 

\noindent is shown. It follows the same steps as the divergence of the continuous flux at the $k$-th cell (see equation \eqref{eqn:DivContFlux}) with the difference that, instead of the numerical upwind flux value recovered at the cell interfaces, the average values between neighbors are used instead,

\be
q^{l,\text{avg}}_{k-\frac{1}{2}} = \frac{q^l_{k-1}(+1) + q^l_k(-1)}{2} \quad \text{and} \quad q^{l,\text{avg}}_{k+\frac{1}{2}} = \frac{q^l_{k}(+1) + q^l_{k+1}(-1)}{2}.
\ee

\noindent Ultimately, the divergence of the SFS closure terms are added to the divergence of their respective fluxes, still performed with the Lax-Friedrichs procedure \eqref{eqn:DivContFlux}, to complete the system of equations and advance it in time. The step of averaging the SFS closure values at the cell interfaces is essential for numerical stability of the scheme because it is able to account for the instances when a discontinuity is traveling between cells. At the moment they are traveling between cells, the complete strength of the discontinuity is unknown in each separate cell and, therefore, the sharing of information at the interface aids the scheme in applying the necessary dissipation magnitude.}
\\

\reviewerA{
\subsection{Inspection of the modulation transfer function curve} \label{subsec:ModTransFunct}
}

\reviewerA{This subsection explains the details of an {\it a priori} analysis based on the exact solution to the Burgers' equation used to inform the design of the modulation transfer function curve \eqref{eq:modulation_transfer_function}.

An {\it a priori} analysis carried out in the context of large-eddy simulations exploits the solution coming from a direct numerical simulation (DNS) of the Navier-Stokes equations, which is taken as a reference in lieu of an unavailable analytical solution. The DNS solution is sharp-spectral-filtered ($\overline{\cdot}$) to compute the exact sub-filter scale (SFS) stresses,  $\tau^{ij} = \overline{u^iu^j} - \overline u^i \overline u^j$ \citep{Clark_JFM_1979,PiomelliMF_PoF_1988,GermanoPMC91}. These exact SFS stress values are compared against the output of SFS models driven by the exact filtered solution, $\overline u^i$, allowing to focus the analysis only on modeling errors.

The same procedure is used in the current manuscript with two important differences: (1) analytical solutions for 1D shock dominated flows are used as a reference solution instead of numerical DNS results; (2) the primary filter operation is sharp in the Legendre (rather than in the Fourier) spectral space. In one dimension the SFS stress tensor is reduced to the single component $\tau^{11} = \overline{u^1u^1} - \overline u^1 \overline u^1$, which is the focus of the following {\it a priori} analysis.} 

\reviewerB{The Burgers' test case, described in more detail in section \ref{subsec:Burgers}, is used in the {\it a priori} analysis shown in figure \ref{fig:aprioriModulation} to showcase the spectral behavior of the exact SFS stress and the effect of the spectral modulation function \eqref{eq:modulation_transfer_function}. It can be observed that the magnitude of the exact $\widetilde{\tau}{^{11}}$ -- the Legendre transform of the exact $\tau^{11}$ term -- generally increase with $\eta$ with a convex shape, culminating with a peak near the cutoff, $\eta = 1$. A similar trend was also observed in theoretical studies of the spectral behavior of sub-filter scale turbulence by \citet{kraichnan1976eddy} and \citet{chollet1981parameterization}. It was observed that if the primary filter cutoff lies in the inertial subrange, the resulting spectral decomposition of the eddy viscosity is not a flat function of the resolved wavenumber space: it exhibits, rather, a plateau at low wavenumbers and a sharp rise near the grid cutoff, referred to as the \emph{plateau-cusp} behavior. A similar behavior was observed in the SFS stress term extracted from the Burgers' problem with a sharp spectral Fourier filter and such connection was used as an argument for building a unified shock capturing and turbulence model, as reported in \citet{Sousa_ArXiv_2021}.

Figure \ref{fig:aprioriModulation} also shows how spectral distribution of the SFS stresses predicted by the \modelAcronym~model without the modulation operation \eqref{eqn:Mod} exhibits the wrong trend. The adoption of unmodulated SFS stresses would lead to increased dissipation values at the larger scales, i.e. lower Legendre modes, and could affect the accuracy of the resolved field. This effect is discussed further in \ref{sec:Appendix}. The results gathered from the \modelAcronym~model with the modulation operation \eqref{eqn:Mod} active, follow much more closely the exact SFS stress component, removing unnecessary dissipation at the larger resolved scales.}
%-------------------------------------------------------------------------------------------------------------------------------------------------------------------------%

%-------------------------------------------------------------------------------------------------------------------------------------------------------------------------%
%!TEX root = ../SousaLSV_JCP_2022.tex

\section{Performance assessment of the \modelAcronym's closure}\label{sec:PerfLSV}

In this section, the proposed \modelAcronym's closure will be tested. First, the developed methodology is applied to the 1D Burgers' equation. Next, problems related to the inviscid compressible Navier-Stokes system of equations, i.e. the Euler equations, are tackled. Initially, the one-dimensional problems of the Riemann shock tube problem \citep{SOD_1978_JCP} and the Shu-Osher shock-entropy wave interaction \citep{shu1988efficient} are addressed. Moving forward, a 2D inviscid strong shock-strong vortex interaction simulation is executed and, ultimately, a simulation of the 2D double Mach reflection problem is carried out. The aforementioned test cases are solved using a \reviewerA{3rd-order strong stability preserving (SSP) Runge-Kutta \citep{gottlieb2001strong}} time integration method.

\subsection{Burgers' Equation}\label{subsec:Burgers}

The inviscid Burgers' equation is a nonlinear scalar conservation \reviewerA{law} that develops a discontinuity in a finite time and can be regarded as a simplified model of the nonlinearity present in the full Navier-Stokes system. Hereafter it is introduced alongside with its filtered version,

\begin{equation}  \label{eqn:FiltBurgers}
\frac{\partial u^1}{\partial t} + \frac{1}{2} \frac{\partial u^1 u^1}{\partial x^1} = 0,  \quad 
 \frac{\partial \overline u^1}{\partial t} + \frac{1}{2} \frac{\partial \overline u^1 \overline u^1}{\partial x^1} =  -  \frac{1}{2} \frac{\partial \tau^{11}}{\partial x^1}.
\end{equation}

\noindent The same procedure used to achieve the filtered compressible Navier-Stokes equations \eqref{eqn:FavreCont} - \eqref{eqn:CoE} is repeated here to get the filtered Burgers' equation. Note that, in this simple equation, only one component ($\tau^{11}$) of the SFS fluxes is present. Its simplicity, though, allows for an analytical solution through the method of characteristics given a certain initial condition and, therefore, allows for a performance assessment of the \modelAcronym\ closure model in both {\it a priori} and {\it a posteriori} analysis. The following initial condition, 

\begin{equation} \label{eqn:BurgersInitCond}
u^1(x,t=0) = 1 + \frac{1}{2}\sin(\pi x),
\end{equation}

\noindent in a periodic domain $x^1 \in [-1,1]$ is chosen as the basis of the numerical study. It represents a traveling nonlinear pulse which steepens until a discontinuity is formed.

\begin{figure}[ht]
\centering
\includegraphics[width=.24\linewidth]{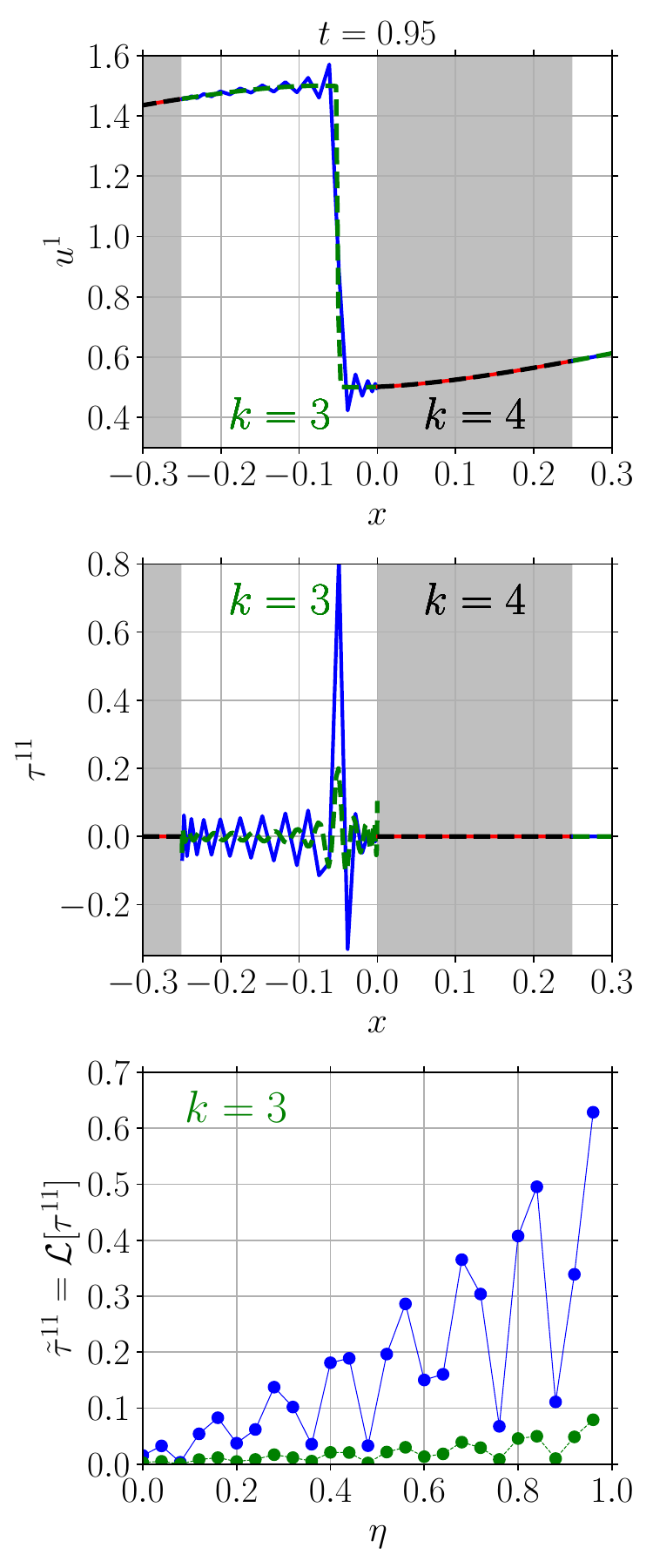}
\includegraphics[width=.24\linewidth]{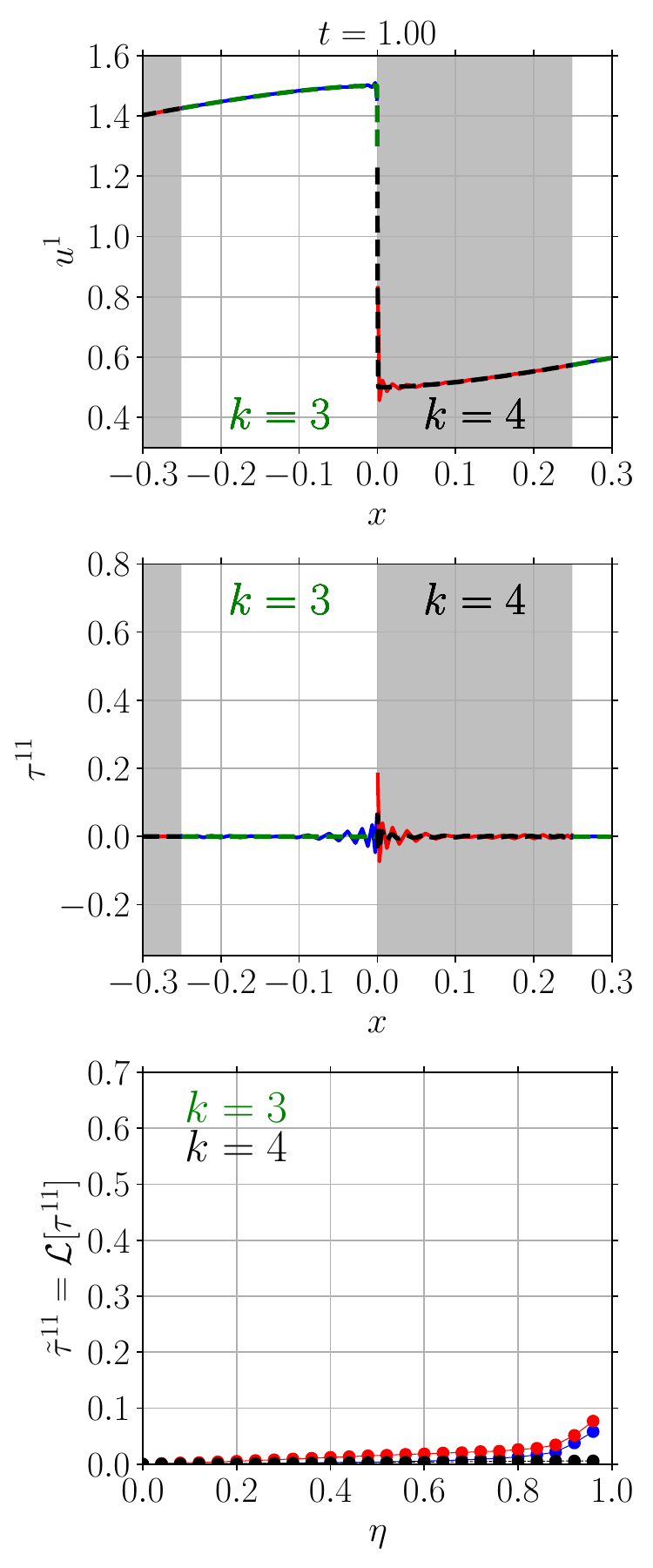}
\includegraphics[width=.24\linewidth]{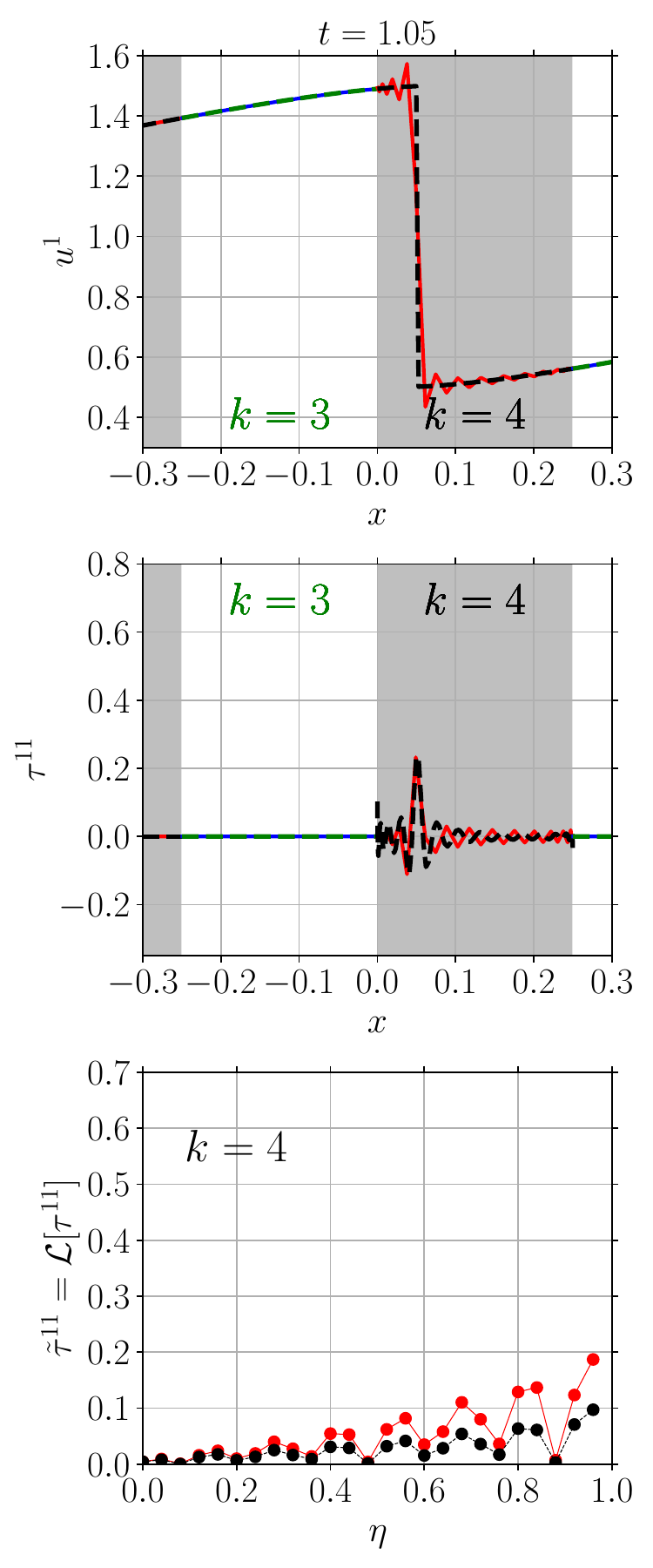}
\includegraphics[width=.24\linewidth]{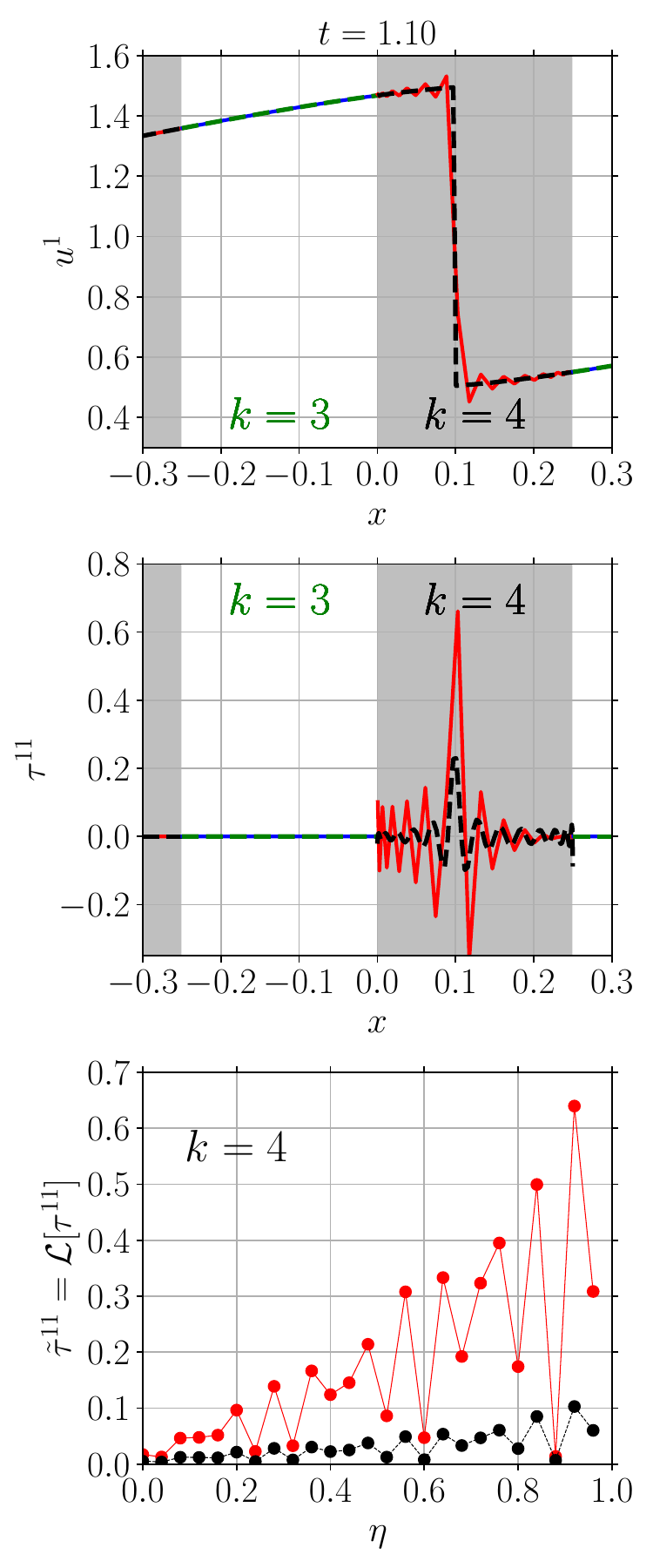}
\caption{{\it A priori} analysis performed on the Burgers' equation with initial conditions given in equation \eqref{eqn:BurgersInitCond} when the periodic domain $x^1 \in [-1,1]$ is discretized with $K = 8$ cells and polynomial order of $N=25$. \reviewerA{Exact sub-filter stress results and exact solution are shown in alternating black or green to showcase the discontinuity between cells. For the same reason, modeled $\tau^{11}$ and filtered solution results are shown in either blue or red.}
}\label{fig:aprioriBurgers} 
\end{figure}

\reviewerA{Figure \ref{fig:aprioriBurgers} gathers results from an {\it a priori} analysis, explained at the beginning of section \ref{subsec:ModTransFunct}, where the periodic domain $x^1 \in [-1,1]$ was split in $K = 8$ cells, each discretized with polynomials of order $N = 25$. It is important to notice that the exact solution, sharp-spectral-filtered in the Legendre space, displays slight overshoots and undershoots near the discontinuity. Although the presence of Gibbs oscillations in the solution might be undesirable in some applications, they are the output of a sharp spectral filter operation, the one which keeps the most information intact for a certain grid resolution. To suppress such oscillations, a large portion of the resolved spectrum would have to be attenuated. 
%That could lead to the spurious dampening of physical high wavenumber oscillations when these are present in the flow field. 
Further discussion is provided in \ref{sec:Appendix}, which analyzes the effects of an increased dissipation across all the resolved wavenumbers in the Sod shock tube and Shu-Osher problems. Ultimately, the \modelAcronym\ closure naturally retains some small-amplitude high-frequency oscillations in exchange for its robust performance and resolving power on relatively coarse grids.}

Furthermore, the results show that both the modeled and exact $\tau^{11}$ are contained within one cell when the discontinuity lies entirely inside it and within at most 2 neighboring cells when the discontinuity lies between the cells' interfaces, i.e. at $t = 1$. If the spectral content of $\tau^{11}$ is analyzed, one will notice that the correct wavenumber modulation is introduced and the order of magnitude of the inserted dissipation is fairly recovered. Some degree of overestimation during the shock's advection through the cells is added, rendering a numerically safe scheme.

\begin{figure}[ht]
\centering
\includegraphics[width=.3\linewidth]{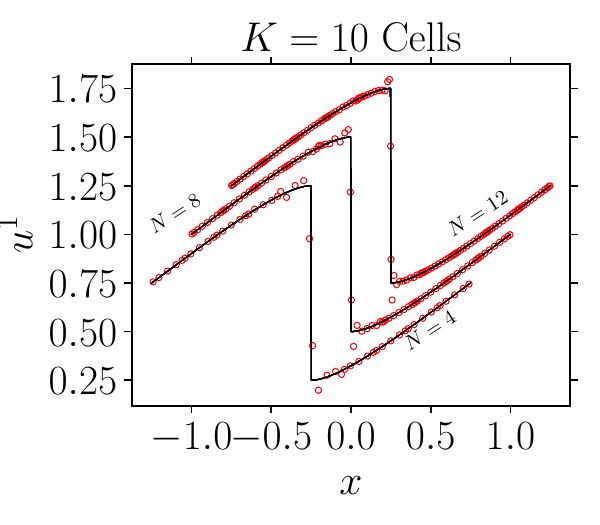}
\includegraphics[width=.3\linewidth]{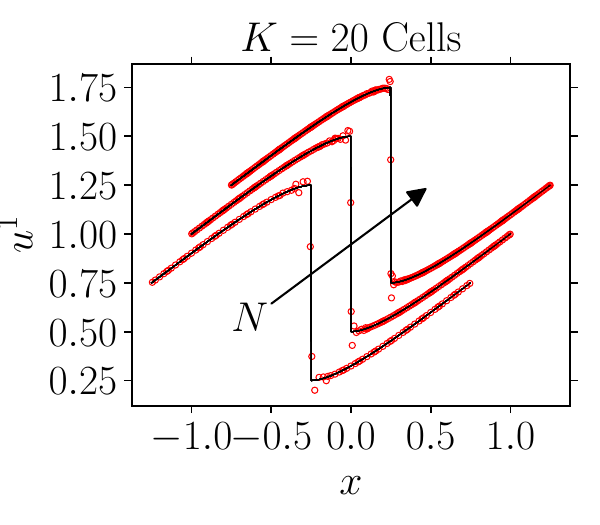}
\includegraphics[width=.3\linewidth]{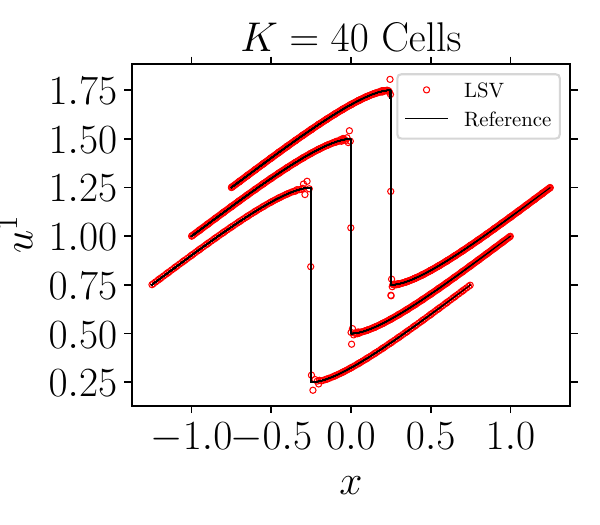}
\caption{High-order flux reconstruction numerical solution of the Burger's equation using the \modelAcronym\ closure. The domain is discretized with a combination of $K = [10,20,40]$ cells and $N=[4,8,12]$ polynomial orders and the initial condition given in equation \eqref{eqn:BurgersInitCond} is solved up to $t = 1$ in a periodic domain $x^1 \in [-1,1]$. The analytical solution is shown in fine solid black lines and the numerical results at the solution points are show in red circular markers.
}\label{fig:aposterioriBurgers} 
\end{figure}

Moving forward, the \modelAcronym\ closure model is used to simulate the evolution of the initial condition, equation \eqref{eqn:BurgersInitCond}, until $t = 1$. The results in figure \ref{fig:aposterioriBurgers} gather the combinations of simulations with $K = 10,\ 20$ and $40$ cells with polynomial orders of $N = 4,\ 8$ and $12$. They show that both the refinement in the number of cells and the expansion in polynomial order are capable of increasing the accuracy of the final solution and concentrating the numerical errors near the region of discontinuity. This is an interesting result that showcases the capability of the model to perform internal cell shock capturing. \reviewerA{As previously mentioned, the Gibbs oscillations near the discontinuities are a consequence of attempting to diminish the dissipation on the resolved spectrum of the solution. }

\begin{figure}[ht]
\centering
\includegraphics[width=.3\linewidth]{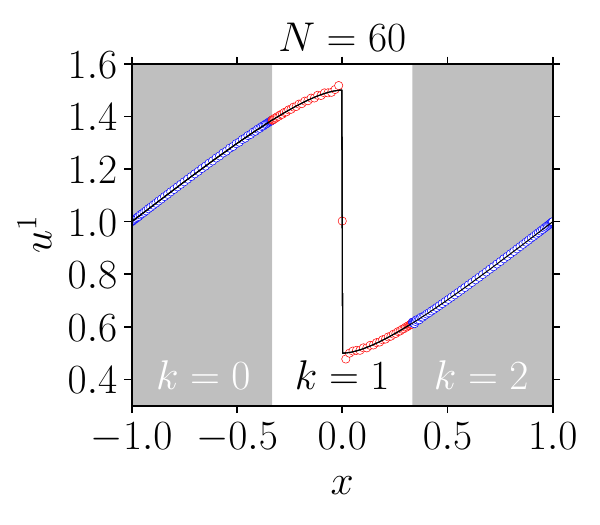}
\includegraphics[width=.3\linewidth]{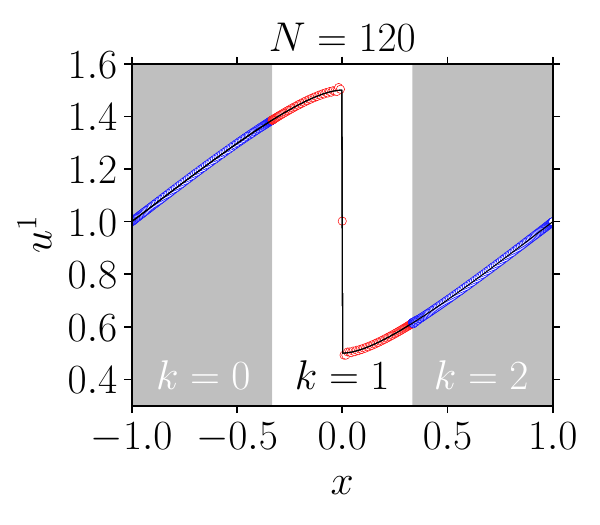}
\includegraphics[width=.3\linewidth]{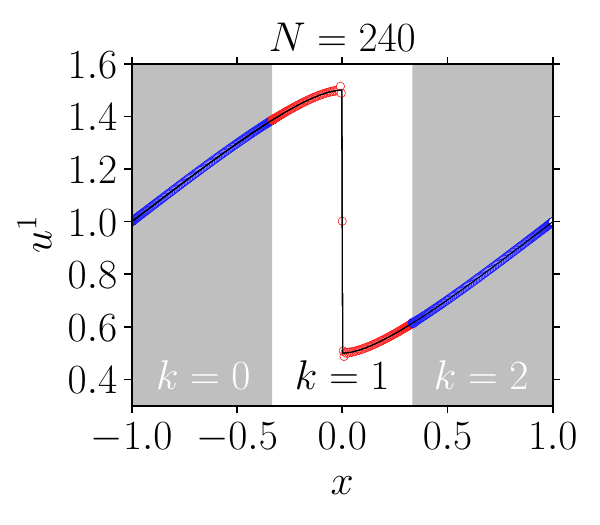}
\caption{Flux reconstruction numerical solution of the Burger's equation using the \modelAcronym\ closure for very high-order polynomials. In this case $K=3$ cells are used to discretize the domain and, in turn, each cell is solved with polynomial orders of $N = [60,120,240]$. The initial condition given in equation \eqref{eqn:BurgersInitCond} is solved up to $t = 1$ in a periodic domain $x^1 \in [-1,1]$. The analytical solution is shown in fine solid black lines and the numerical results at the solution points are show in alternating red or blue circular markers. Regions corresponding to different cells are periodically shaded.
}\label{fig:aposterioriBurgersHigh} 
\end{figure}

Additionally, the current method is shown to be robust at even very high-order of polynomial reconstruction. Inspired by \citet{Asthana_JCP_2015}'s result of a steady Burger's equation solution being captured by a $119$-th order polynomial reconstruction within the center cell of a three-element configuration, figure \ref{fig:aposterioriBurgersHigh} shows results for an advecting nonlinear pulse solved with $K=3$ and $N=60,120$ and $240$. This shows that, not only the current method is capable of performing a very high-order solution of a discontinuity within a cell, but also to allow the discontinuity to move seamlessly from a neighboring cell to another.

\subsection{Sod shock tube}\label{subsec:Sod}

Moving forward, the focus is changed from a scalar nonlinear conservation law to the inviscid Navier-Stokes system of equations in one dimension, i.e. the Euler equations. If the aforementioned system of equations is initialized with the following initial conditions in density, velocity and pressure,

\begin{equation} 
[\rho,u^1,p](x,0) = 
\begin{cases}
    [1,0,1],& \text{if } x <  0.0\\
     [0.125,0,0.1],              & \text{otherwise},
\end{cases}
\end{equation}

\noindent for $x^1 \in [-1,1]$, then a shockwave propagating to the right develops. An exact solution can be found for this setting, known as the Sod shock tube problem \citep{SOD_1978_JCP}, and it consists of 4 constant density states separated by a shock, a contact discontinuity and a rarefaction wave.

\begin{figure}[ht]
\centering
\includegraphics[width=.9\linewidth]{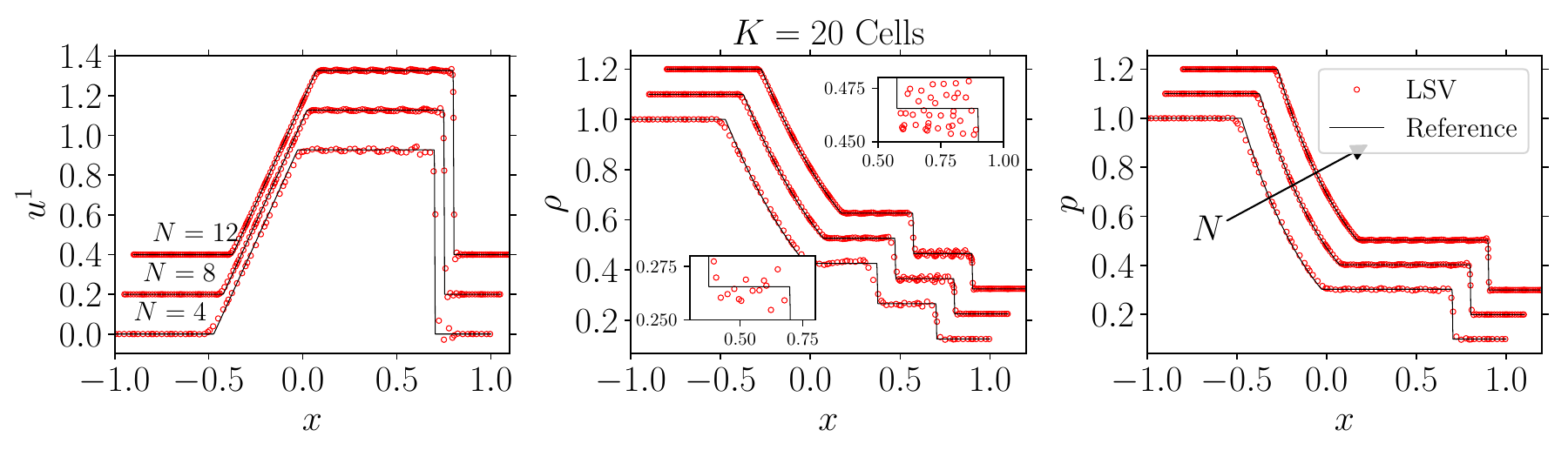}
\includegraphics[width=.9\linewidth]{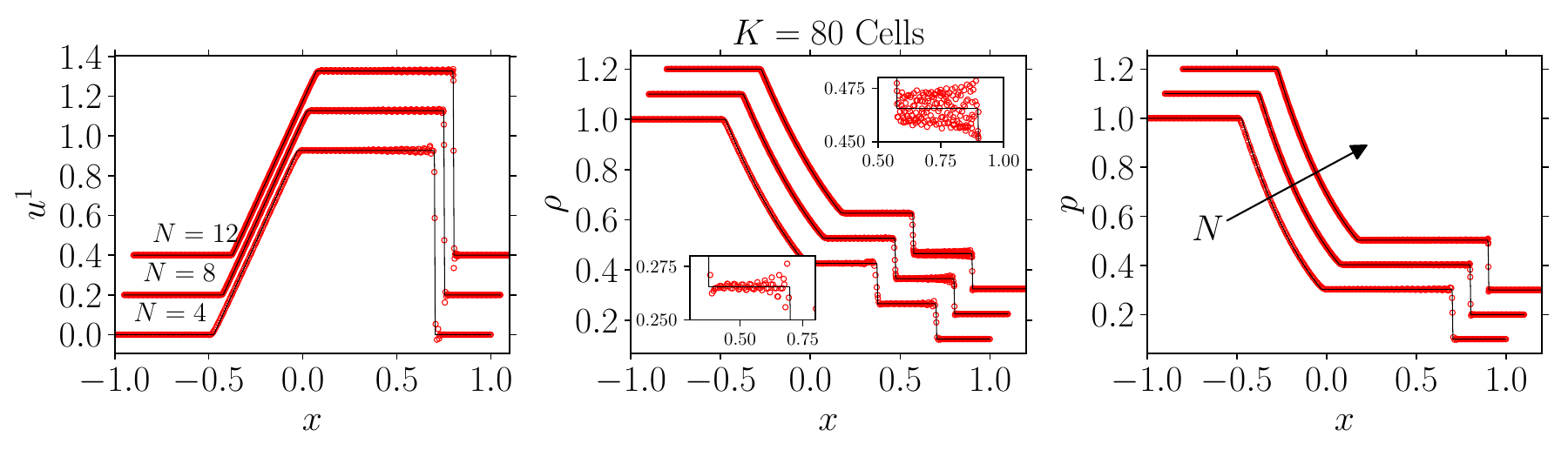}
\caption{High-order flux reconstruction numerical solution of the Sod shock tube problem \citep{SOD_1978_JCP} using the \modelAcronym\ closure. The domain is discretized with a combination of $K = [20,80]$ cells and $N=[4,8,12]$ polynomial orders and the initial condition is solved up to $t = 0.4$. The analytical solution is shown in fine solid black lines and the numerical results at the solution points are show in red circular markers.
}\label{fig:aposterioriSOD} 
\end{figure}

In progression, the Sod shock tube problem \citep{SOD_1978_JCP} is simulated up to $t=0.4$. The computational domain is discretized with combinations of $K = [20,80]$ cells and polynomials of order $N=[4,8,12]$. Figure \ref{fig:aposterioriSOD} gathers the results of this numerical experiment. It can be observed that a refinement in either the total number of cells or the order of the polynomial reconstruction within each cell leads to a sharper gradient capturing near discontinuities as well all as increased accuracy on the remaining of the computational domain. Throughout the domain, but mainly in the region between the contact discontinuity and the shock in the density field, it is possible to observe some high wavenumber spurious oscillations for all the number of cells and polynomial orders tested, also present in simulations performed by \citet{HagaKawai_JCP_2019}. These oscillations are low in amplitude, at most $\delta \rho = 0.03$, and are more pronounced at higher-orders. Their amplitude decreases with cell number refinement but superior levels of refinement are needed if higher order polynomials are used. \reviewerA{Another strategy is to further increase the dissipation levels of the resolved spectrum but this may lead to unnecessary damping of the physically relevant oscillations in the resolved field that occur near the resolution limit. An analysis of such a scenario is provided in \ref{sec:Appendix}.} 

\begin{figure}[ht]
\centering
\includegraphics[width=.9\linewidth]{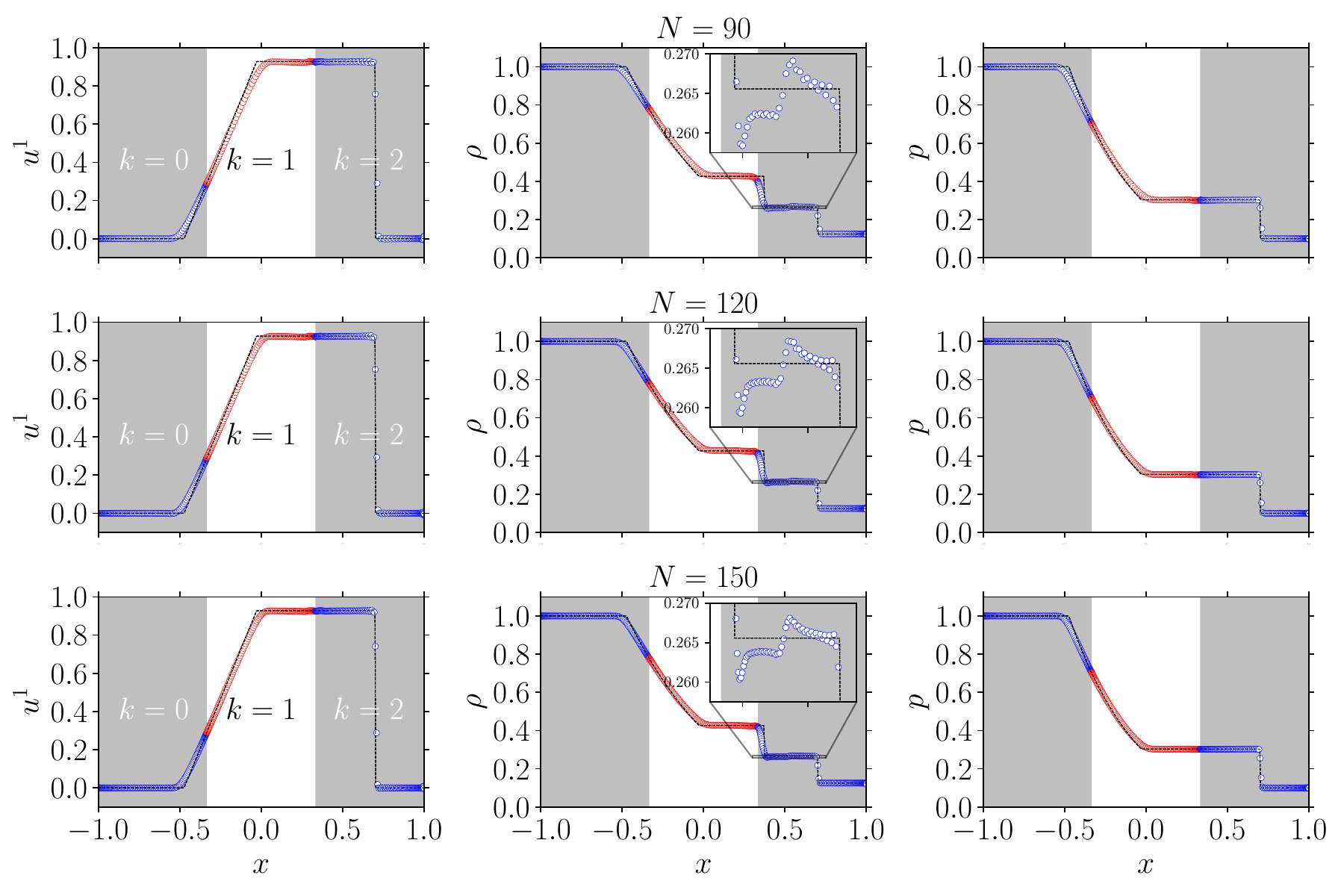}
\caption{\reviewerA{Flux reconstruction numerical solution of the Sod shock tube problem \citep{SOD_1978_JCP} using the \modelAcronym\ closure for very high-order polynomials. In this case $K=3$ cells are used to discretize the domain and, in turn, each cell is solved with polynomial orders of $N = [90,120, 150]$. The analytical solution is shown in fine solid black lines and the numerical results at the solution points are show in alternating red or blue circular markers. Regions corresponding to different cells are periodically shaded.}
}\label{fig:aposterioriSODHigh} 
\end{figure}

Once more, the robustness of the current method is shown, in figure \ref{fig:aposterioriSODHigh}, by simulating a Sod shock tube problem, a nonlinear coupled system, using polynomials orders of $N = [90,120, 150]$ with $K = 3$ cells. This is an extension of what was previously published by \citet{Asthana_JCP_2015}, who limited their very high-order simulations to the Burgers' equation. Moreover, although the simulation is performed at $N \ge 90$, the presence of spurious oscillations is greatly attenuated. This is a consequence of the domain being discretized with only three separate cells. Because of that, at least a third of the domain, i.e. the cell in which the discontinuity is present, is subjected to the dissipation added by the \modelAcronym~closure. This can be desirable if no physical small scales are present in the simulation but it can also lead to a decrease in resolution power if that is not the case, \reviewerA{as observed in the Shu-Osher problem (see figure \ref{fig:aposterioriShuOsherDOF}). Despite a decrease in the presence of low-amplitude high-frequency oscillations, a small kink of amplitude $ \delta \rho \approx 0.01$ is observed in the density field between the contact discontinuity and the main shock. This is a consequence of allowing small overshoots to occur in exchange for the ability to retain high wavenumber information. Ultimately, this is not a concern since the amplitude of the kink decreases with an increase in the polynomial order, as the method converges to the exact solution.} 

\subsection{Shu-Osher shock-entropy wave interaction} \label{subsec:ShOsher}

Moving forward, focus is given to the shock-entropy wave interaction introduced by \citet{shu1988efficient}. The setup comprises a shock wave propagating into a resting fluid where small amplitude density perturbations are present. Upon interaction with the shock, two phenomena occur: the initially smooth density perturbations steepen and form weak shocks; small-scale acoustic waves that trail the main shock are generated. This numerical test case is challenging because sufficient dissipation is needed to capture the shock discontinuity but, at the same time, the overall dissipation must be limited to prevent a spurious attenuation of the trailing waves.

\begin{figure}[ht]
\centering
\includegraphics[width=.3\linewidth]{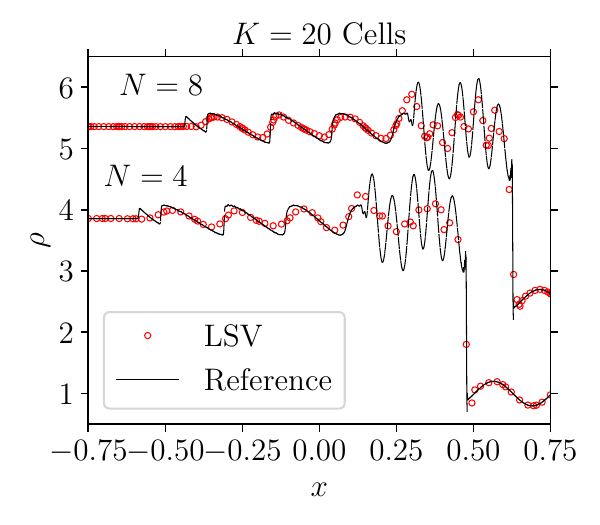}
\includegraphics[width=.3\linewidth]{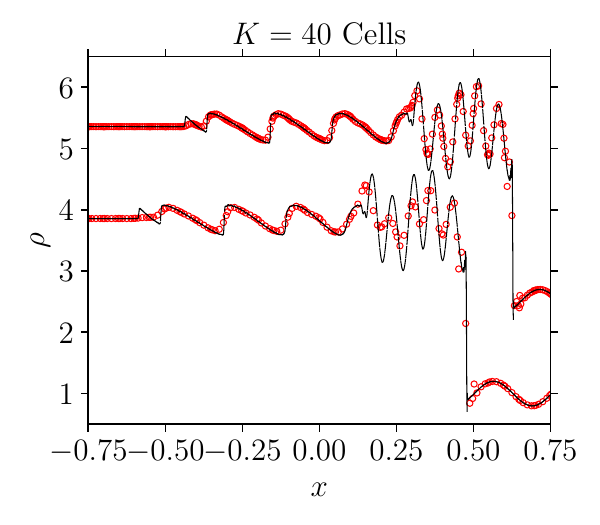}
\includegraphics[width=.3\linewidth]{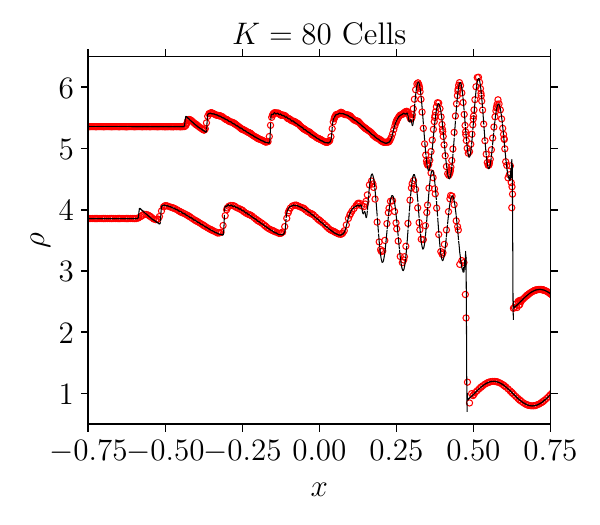}
\caption{High-order flux reconstruction numerical solution of the Shu-Osher shock-entropy wave interaction  \citep{shu1988efficient} using the \modelAcronym\ closure. The domain is discretized with a combination of $K = [20,40,80]$ cells and $N=[4,8]$ polynomial orders and the initial condition is solved up to $t = 0.36$. Numerical results at the solution points are show in red circular markers and are compared against a DNS performed with 500 cells and order 3, shown in fine solid black lines.
}\label{fig:aposterioriShuOsher} 
\end{figure}

 The Shu-Osher shock-entropy wave interaction problem is defined by the following initial conditions in the domain $x \in [-1,1]$,

\begin{equation} 
[\rho,u,p](x,0) = 
\begin{cases}
    [3.857143, 2.629369, 10.333333],& \text{if } x <  -0.8\\
     [1.0 + 0.2\sin(25 x), 0.0, 1.0],              & \text{otherwise},
\end{cases}
\end{equation}

\noindent which are advanced up until $t = 0.36$. Figure \ref{fig:aposterioriShuOsher} gathers the results for the density field when the \modelAcronym\ closure is used and the computational domain is discretized with a combination of $K = [20,40,80]$ cells and a polynomial order equivalent to $N=[4,8]$. The outcome of such simulations are compared against a finely discretized run with 500 cells and order 3, taken here as a direct numerical simulation of the Shu-Osher problem. It can be observed that the refinement in either direction regarding total number of cells or inner cell polynomial order lead to better results in respect to sharper gradients near discontinuities and smaller attenuation of the trailing wave packet. A subset of these results can be compared against similar simulations performed by \citet{HagaKawai_JCP_2019}, where 400 degrees of freedom (DOF) where used. The case where $K=40$, $N=8$ and  $K=80$, $N=4$ have 360 and 400 DOFs respectively and both show qualitatively similar results when compared to \citet{HagaKawai_JCP_2019}'s. Despite the similarity in resolution capacity, the current method is able to surpass the 4-th order polynomial order limitation existent in the aforementioned method.

\begin{figure}[ht]
\centering
\includegraphics[width=.9\linewidth]{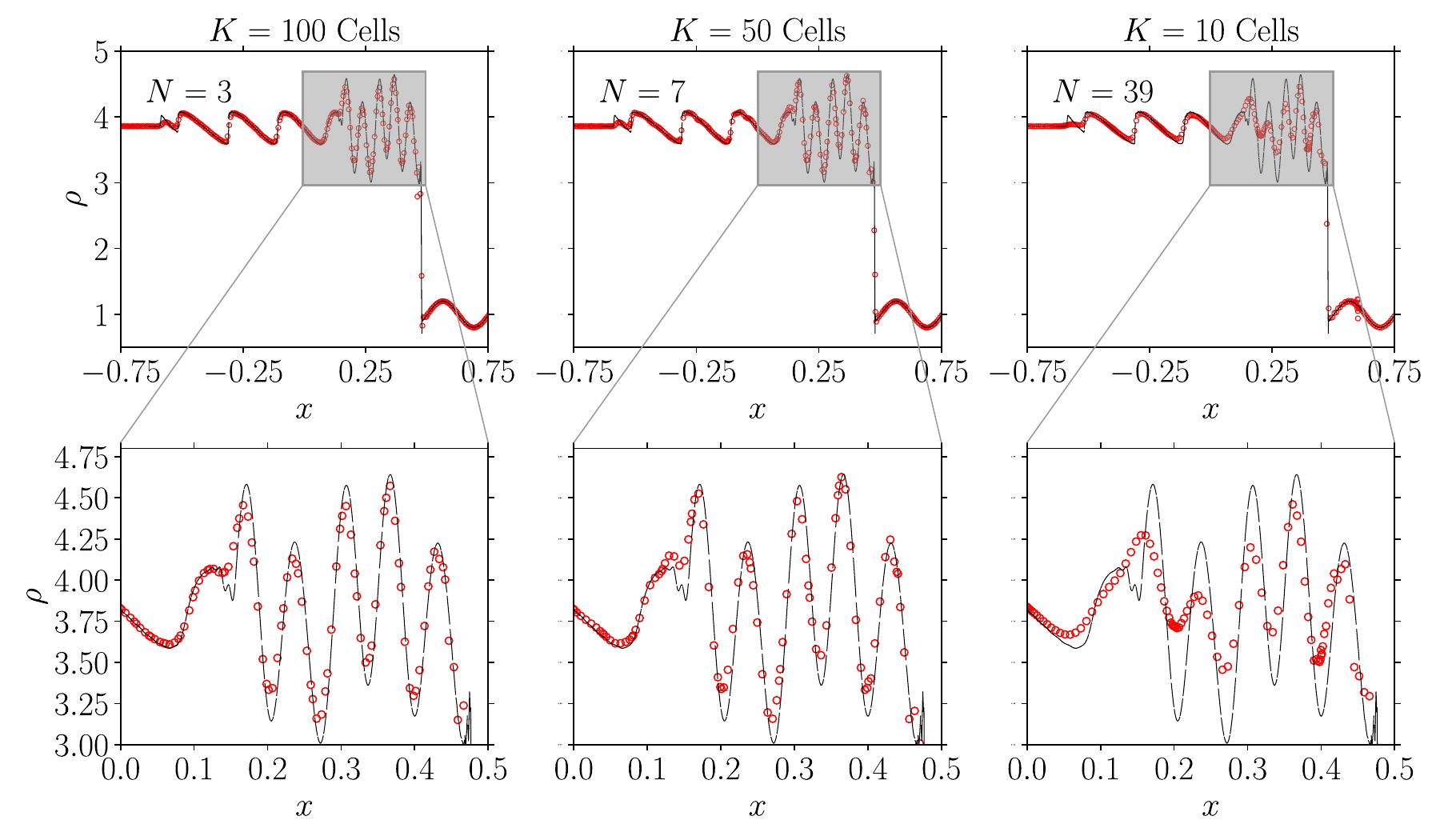}
\caption{High-order flux reconstruction numerical solution of the Shu-Osher shock-entropy wave interaction  \citep{shu1988efficient} using the \modelAcronym\ closure. The domain is discretized with 400 DOF and the initial condition is solved up to $t = 0.36$. Numerical results at the solution points are show in red circular markers and are compared against a DNS performed with 500 cells and order 3, shown in fine solid black lines.
}\label{fig:aposterioriShuOsherDOF} 
\end{figure}

Now, if the number of degrees of freedom (DOF) is kept constant but different simulations are performed varying the number of cells and polynomial order accordingly, an important limitation of the model is made evident. \reviewerB{A detailed analysis of figure \ref{fig:aposterioriShuOsherDOF} shows that, although the solution quality improves for the same number of DOF with an initial increase in polynomial order from $N=3$ to $N=7$, it deteriorates if further higher orders are considered. This effect is explained by the constant DOF constraint, as an increase in polynomial order is accompanied by a decrease in cell numbers and, hence, larger cell sizes. Since the \modelAcronym~model is able to dynamically adjust the magnitude of the added dissipation by extracting  information from the resolved kinetic energy near the cutoff, it is mainly active in the cell which contains the primary shock discontinuity. Larger cells, therefore, spread out the dissipation over a wider area surrounding the main shock, rendering the scheme overall more dissipative. In conclusion, with the current formulation, a high level of accuracy can be recovered if sufficiently small cell sizes are used. If that is the case, an increase in polynomial order can increase the solution quality. The presence of relatively large cells, though, can decrease the model's performance. }

This drawback can be circumvented by using a strategy similar to the semi-local energy at the cutoff estimation reported in \citet{Sousa_ArXiv_2021}. The strategy is based on the connection between the spectral and physical spaces which render local quantities in one space global in the other and vice-versa. Ultimately this is just a property of convolution integrals and is the same reason behind Heisenberg's uncertainty principle. Moving on, if one recovers the exact value of the energy at the cutoff, although accurate, this value will be global in physical space and will be active in the whole cell where nonlinear interactions are important. If this operation is observed in the point of view of a convolution, the use of a delta function as a convolution kernel to pick out the last mode's energy is responsible for the global behavior in physical space. A semi-local energy can be recovered if the delta function kernel is made wider. This effect can be achieved by performing a Legendre filtering operation on the kinetic energy field, for example,

\be \label{eqn:semilocalKE}
 E^1_{N}\left(\overline u^1 \right) \approx \left|{\mathcal{M}}\left[\frac{e^{4\eta}- 1}{e^4 - 1}, \frac{1}{2}(\overline u^1)^2\right]\right|.
\ee

\begin{figure}[ht]
\centering
\includegraphics[width=.24\linewidth]{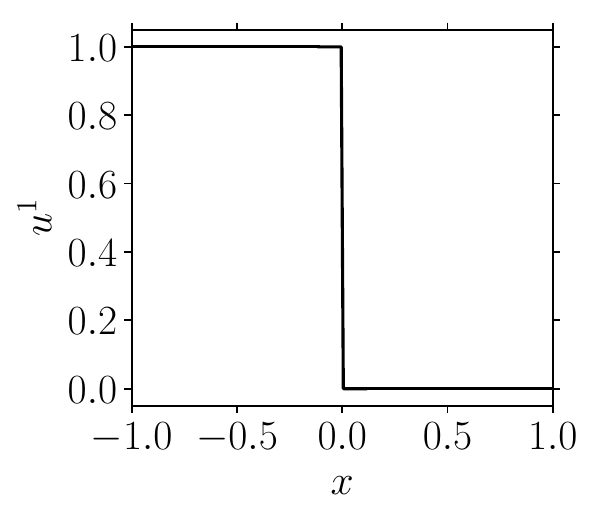}
\includegraphics[width=.24\linewidth]{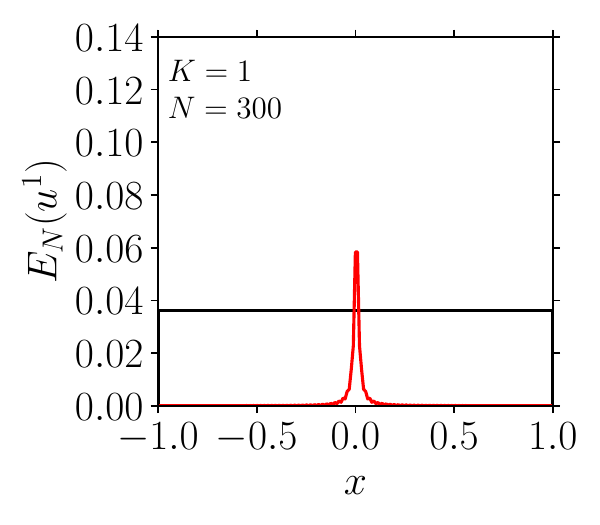}
\includegraphics[width=.24\linewidth]{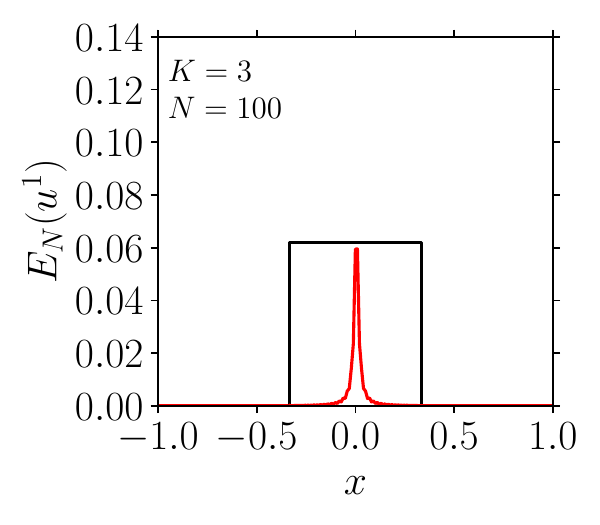}
\includegraphics[width=.24\linewidth]{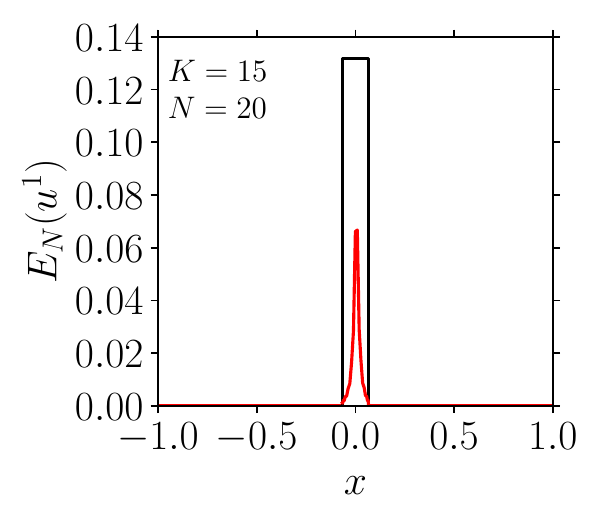}
\caption{Representation of the spectral \eqref{eqn:spectralKE} and semi-local \eqref{eqn:semilocalKE} cutoff energy estimation methods operating on a unitary jump function centered at $x = 0$, shown in the left. On the right, the results for the energy at the cutoff in the physical space is shown for different discretization strategies with respect to total cell number and polynomial order. In black, the spectral cutoff energy is shown, and, in red, its semi-local counterpart.
}\label{fig:semiLocalE} 
\end{figure}

Figure \ref{fig:semiLocalE} shows how the energy near the resolution limit of a unitary jump function centered at $x = 0$ is recovered by the spectral relation, equation \eqref{eqn:spectralKE}, and by the semi-local method, equation \eqref{eqn:semilocalKE}. It can be observed that although the magnitude of the maximum semi-local cutoff energy is in the same order of magnitude of its spectrally recovered counterpart, it can either overestimate or underestimate the theoretical value depending on the order of polynomial reconstruction used. Additionally, it is possible to infer that the semi-local method is more useful for higher order polynomials.  

\begin{figure}[ht]
\centering
\includegraphics[width=.35\linewidth]{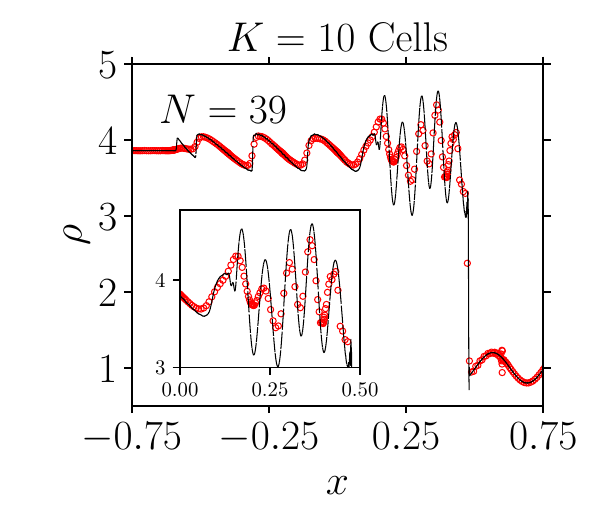}
\put(-120,90){Spectral}
\includegraphics[width=.35\linewidth]{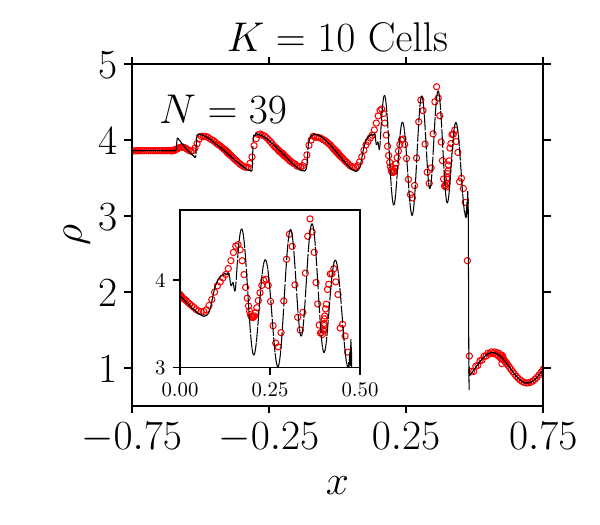}
\put(-124,90){Semi-local}
\caption{High-order flux reconstruction numerical solution of the Shu-Osher shock-entropy wave interaction \citep{shu1988efficient} using the \modelAcronym\ closure. The domain is discretized with $K=10$ cells and a $39^{th}$ order polynomial. The energy at the cutoff is estimated via both the spectral method \eqref{eqn:spectralKE}, shown on the left, and the semi-local method \eqref{eqn:semilocalKE}, shown on the right. The initial condition is evolved up to $t = 0.36$. Numerical results at the solution points are show in red circular markers and are compared against a DNS performed with 500 cells and order 3, shown in fine solid black lines.
}\label{fig:spectralVSsemilocalKE} 
\end{figure}

For example, if the semi-local method is used in the cutoff energy estimation step for the simulation of the Shu-Osher problem with 400 DOF and 10 cells an improvement in the solution's accuracy is achieved in comparison with the default spectral method, see figure \ref{fig:spectralVSsemilocalKE}. It is important to state that the choice of kernel function for the semi-local energy estimation step is not unique. A choice of more localized kernels in spectral space can improve the magnitude estimative but make the physical space counterpart more global and, therefore, a compromise must be made if this path is chosen. Ultimately, the spectral energy estimation method is more convenient, being faster and accurate if the locality constraint is imposed by relatively small cells.

The last item to be addressed in the current subsection is a comparison of results obtained by the current method and the one published by \citet{Asthana_JCP_2015}. For that, a slight modification to the Shu-Osher problem must be made to allow for direct comparison. \citet{Asthana_JCP_2015} performed a change of reference and simulated a stationary shock towards which the initially sinusoidal entropy waves traveled and interacted. Additionally, higher frequency entropy waves are used in the initial condition. Such version of the problem is defined for $x \in [-1,1]$ as 

\begin{equation} 
[\rho,u,p](x,0) = 
\begin{cases}
    [3.857143, -0.920279, 10.333333],& \text{if } x <  0.0\\
     [1.0 + 0.2\sin(50 x), -3.549648, 1.0],              & \text{otherwise}.
\end{cases}
\end{equation}

\begin{figure}[ht]
\centering
\includegraphics[width=.45\linewidth]{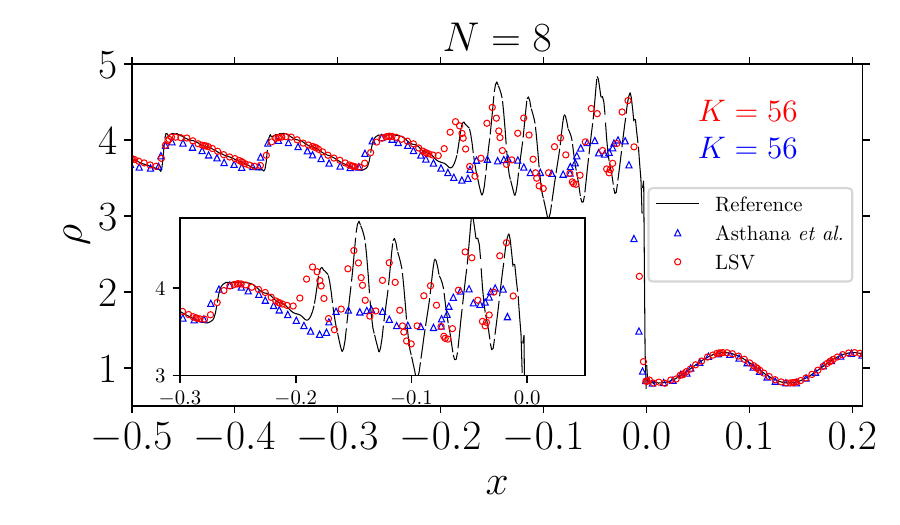}
\includegraphics[width=.45\linewidth]{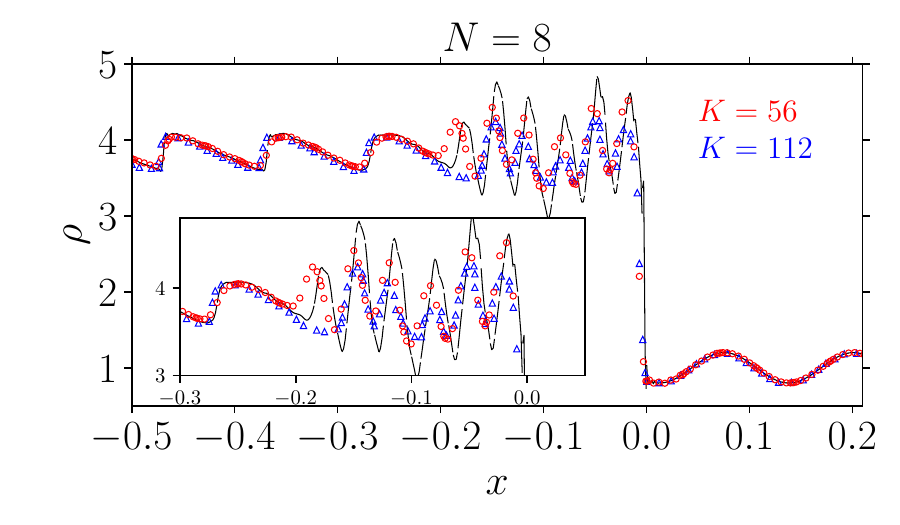}\\
\includegraphics[width=.45\linewidth]{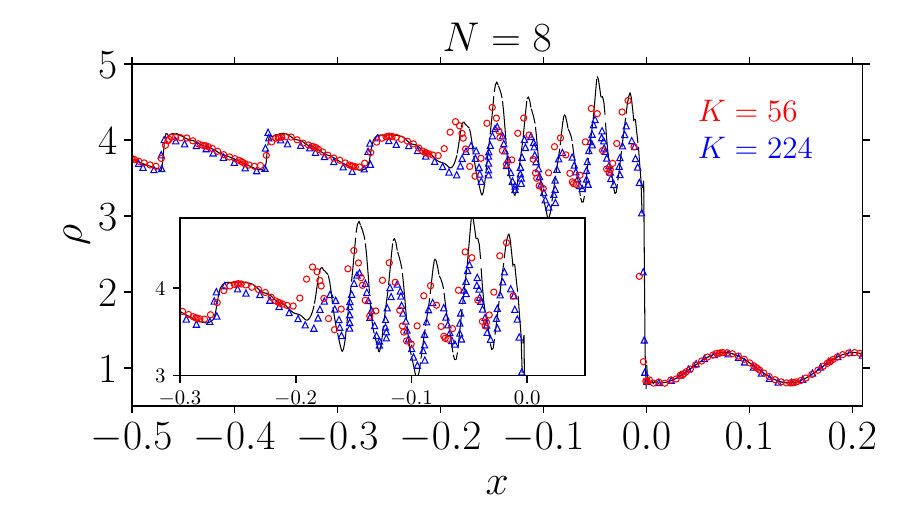}
\includegraphics[width=.45\linewidth]{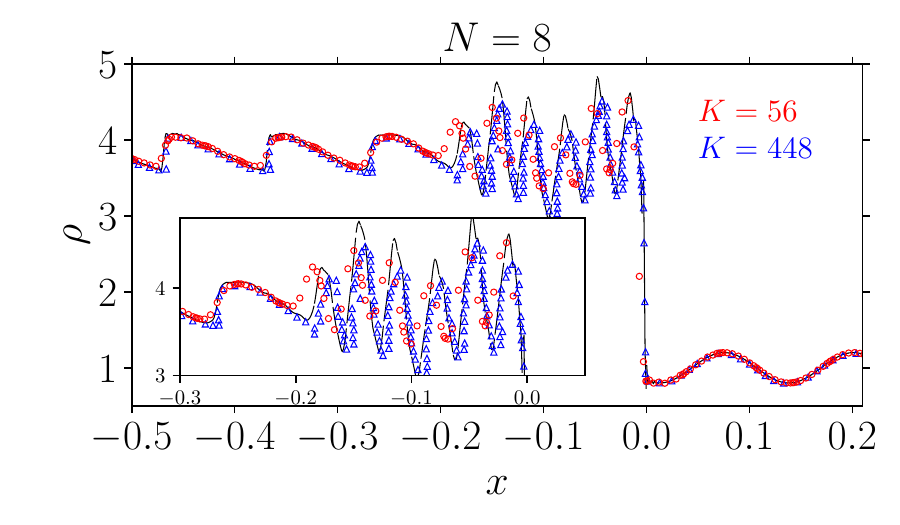}
\caption{High-order flux reconstruction numerical solution of the Shu-Osher Shock-Entropy Wave Interaction  \citep{shu1988efficient} using the \modelAcronym\ closure is compared against \citet{Asthana_JCP_2015}'s method. The domain is discretized with a polynomial order of $N=8$. The results simulated with \modelAcronym\ using $K=56$ cells, shown in red circles, are compared against digitized results published by \citet{Asthana_JCP_2015} for $K=[56,112,224,448]$, shown in blue triangles. A reference solution is obtained by solving the initial condition up to $t = 0.2$ with 500 cells using order 3, shown in fine dashed lines.
}\label{fig:aposterioriShuOsherAsthana} 
\end{figure}

The comparison between the current method based on the \modelAcronym\ closure and \citet{Asthana_JCP_2015}'s results is displayed in figure \ref{fig:aposterioriShuOsherAsthana}. The present implementation outperforms the previously published method with respect to the capacity of accurately solving the small scales that occur in the wavepacket trailing the main shock discontinuity. This conclusion is drawn because, even though the same polynomial order of $N=8$ was used for both methods, if the same number of cells of $K=56$ is used, the \modelAcronym\ based simulation is able to recover most of the rear shock dynamics and, in turn, \citet{Asthana_JCP_2015}'s method dissipates much of the high wavenumber oscillations. In fact, the previously published method needs approximately 8 times as much cells, $K=448$, to retain a similar accuracy in the post shock region. \reviewerA{It is possible that the parameters for \citet{Asthana_JCP_2015}'s method might be able to be better adjusted and optimized to resolve the flow physical small-scale content. That being said, the version of the model reported in the literature has shown difficulties in resolving such waves.}

\subsection{Inviscid strong shock-strong vortex interaction} \label{subsec:shockVortex}

A stationary shock supported by a supersonic inflow velocity of $V_0 = 1.5\sqrt{\gamma p_0/\rho_0}$ is initialized at $x_s/L = 1/2$ inside a computational domain $\Omega = [0,2L]\times[0,L]$. A compressible zero-circulation vortex is initialized upstream of the shock at $(x_v/L,y_v/L) = (1/4,1/2)$ with an inner core radius equal to $a/L = 0.075$ and an external radius $b/L = 0.175$. This can be translated as ${\bf u } = u_{\theta}(r)  {\bf \hat e}_{\theta} + V_0  {\bf \hat e}_{x}$, where $r$ is the radial distance from the center of the vortex and 

\begin{equation}
 \frac{u_{\theta}(r)}{u_{\theta}(a)} = 
 \begin{cases}
 \frac{r}{a} ,& \text{if } r \leq  a\\
\frac{\eta}{2}\left( \frac{r}{b} -  \frac{b}{r}\right)  ,& \text{if } a < r  \leq b\\
 0,& \text{otherwise,}
 \end{cases}
\end{equation}

\noindent where $\eta = 2(b/a)/[1 - (b/a)^2]$ and the maximum tangential velocity is set to $u_{\theta}(a) = 0.9V_0$. Following the pressure field is initialized so that its gradient balances the centripetal force and the following system of equations is solved based on the ideal gas relation and isentropic compression,

\begin{equation}
\frac{\partial P}{\partial r} = \rho \frac{{u}_{\theta}^2(r)}{r}, \quad P = \rho R T, \quad \frac{P}{P_0} = \left(\frac{\rho}{\rho_0}\right)^{\gamma}.
\end{equation}

 \citet{Ellzey_PoF_1995} first employed such a setup to examine the acoustic field generated by the shock-vortex interaction. Following publications by \citet{Rault_JSC_2003} and by \citet{Tonicello_CandF_2020} also used the same arrangement to analyze the driving mechanisms for the production of vorticity due to the interaction with shocks and to investigate shock capturing techniques in high-order methods focused on their influence on the entropy field and its non monotonic profile across a shock, respectively.

\begin{figure}[ht]
\centering
\includegraphics[width=.9\linewidth]{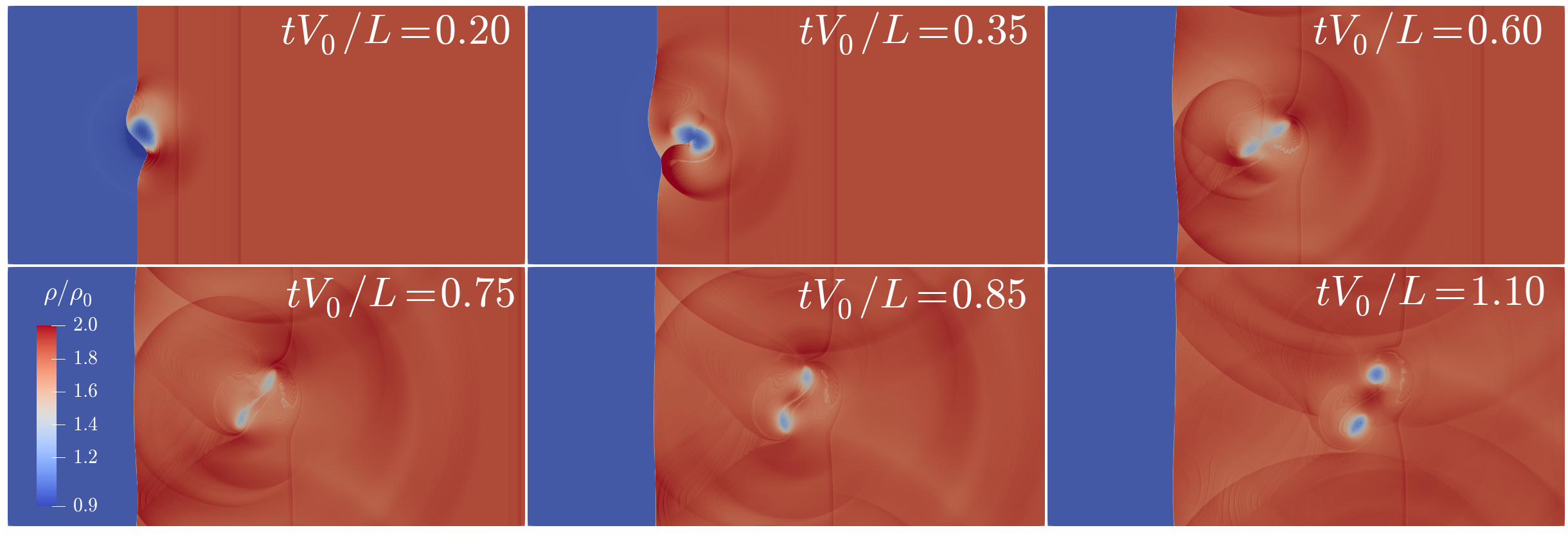}
\caption{\reviewerB{LSV simulation of an inviscid strong shock-strong vortex interaction performed with $N=7$ polynomial order in both directions and $K_x \times  K_y= [256\times128]$ cells, a total of $2097152$ DOF. Six different time instants are shown to illustrate the system's development.
}}\label{fig:ShockVortexPhys} 
\end{figure}

\reviewerB{Previous literature reports that a shock/vortex interaction leads to the compression of an initial circular vortex into an elliptical shape with the ultimate distortion being dependent on shock strength \citep{dosanjh_ShockVortex_1965,Grasso_ShockVortex_2000}. Additionally, for the strong shock, $M_s = 1.5$, and strong vortex, $M_v = 0.9$, case considered, highly resolved simulations show a S-shaped deformation of the shock front upon interaction and, ultimately, a symmetry break which culminates with the formation of two new counter-clockwise rotating vortices, being the top one ahead of its bottom counterpart \citep{Rault_JSC_2003,Tonicello_CandF_2020,Sousa_ArXiv_2021}. A grid-refined LSV simulation (figure \ref{fig:ShockVortexPhys}) displays the aforementioned phenomena as well as the formation of reflected shocks and a Mach stem as discussed by \citet{Ellzey_PoF_1995}, seen around $tV_0/L \approx 0.35$. These results, gathered from a run with approximately 2.1 million degrees-of-freedom (DOF), are also used as reference to compare the resolution power of different grid refinement levels hereafter. }

\begin{figure}[ht]
\centering
\includegraphics[width=.85\linewidth]{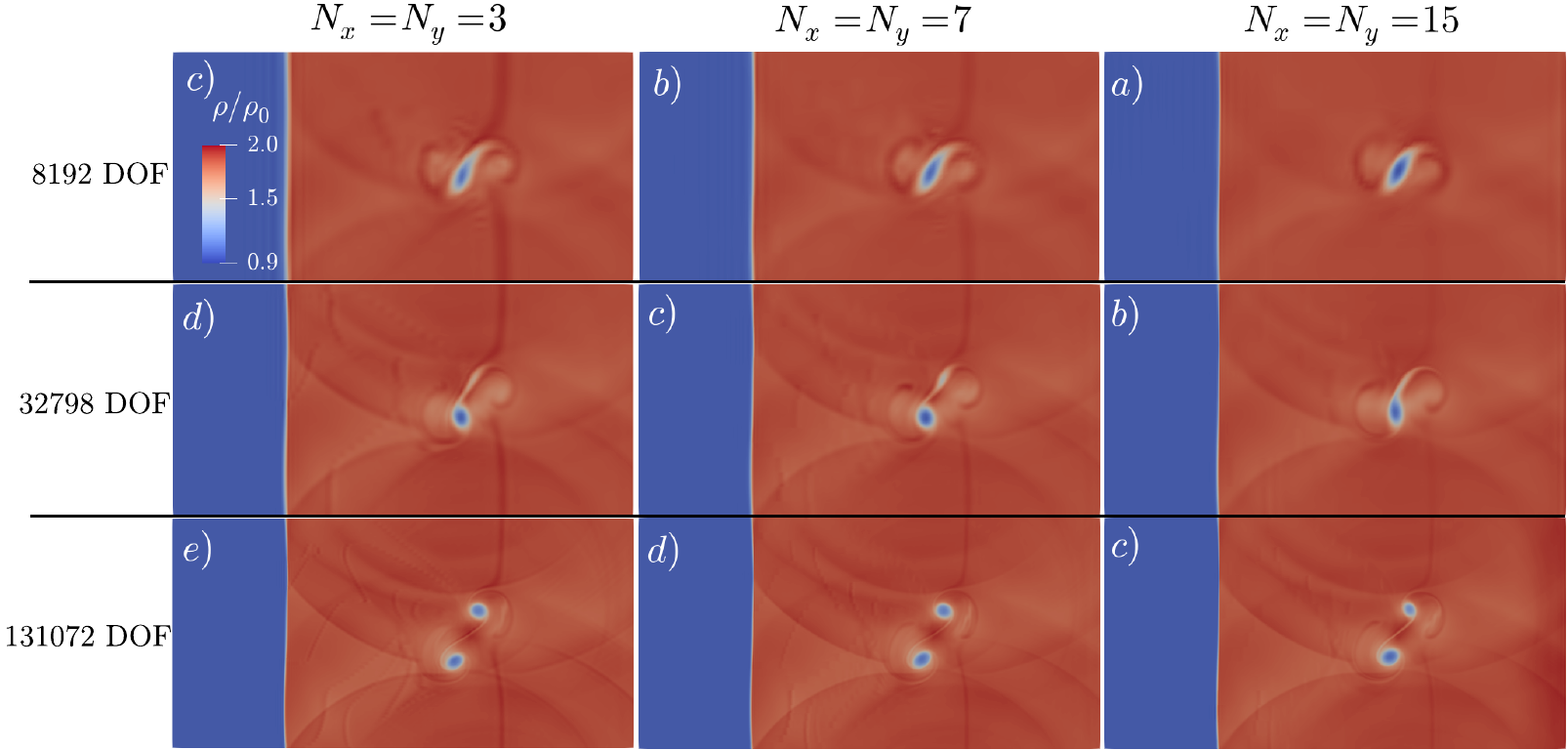}
\caption{\reviewerB{Results for simulations of a strong shock-strong vortex interaction at $tV_0/L = 1.10$ at different polynomial orders, separated by columns, and number of cells, indicated by the letters on the top-left of each subfigure. The cell arrangements used, $K_x \times  K_y$, are: $a$) $8\times4$, $b$) $16\times8$, $c$) $32\times16$, $d$) $64\times32$, $e$) $128\times64$. The combination of the number of cells and and points inside each cell, equal to $N+1$, lead to the total number of degrees-of-freedom (DOF), a measure of the grid refinement level.}
}\label{fig:ShockVortexDOF} 
\end{figure}

\reviewerB{A grid sensitivity analysis is performed for the LSV simulations of such shock/vortex interaction and the results are collected in figure \ref{fig:ShockVortexDOF}. Simulations with different degrees-of-freedom (DOF) were performed and their results are compared at time $tV_0/L = 1.10$, when vortex separation is expected. Initially, it can be inferred that an increase in the number of DOF is directly correlated with an increase in solution quality independently of how the refinement was achieved -- whether with an increase in the number of cells or an increase in the polynomial order used within each cell. Ultimately, at the finest refinement level in this analysis, the solution for all polynomial orders contemplated converged to the reference result.

A more thorough analysis shows that the solution quality for the same DOF first increases and then decays with increasing polynomial order. This is especially evident at the intermediate grid refinement level, where the $N=7$ order simulation is barely able to lead to vortex separation while $N=3$ and $N=15$ are, in this order, further from the expected solution. These results display the same trend illustrated by the investigation of the constant DOF Shu-Osher simulations, shown in figure \ref{fig:aposterioriShuOsherDOF}, and it happens due to the larger cell sizes, associated with higher-order simulations in constant DOF settings, spreading the added dissipation throughout a larger area around the main shock discontinuity. Being the problem similar, the solution is also analogous and a semi-local energy at the cutoff estimation method can also be able to increase the resolution power when larger cells are considered, as observed in figure \ref{fig:spectralVSsemilocalKE}.}

\begin{figure}[ht]
\centering
\includegraphics[width=.85\linewidth]{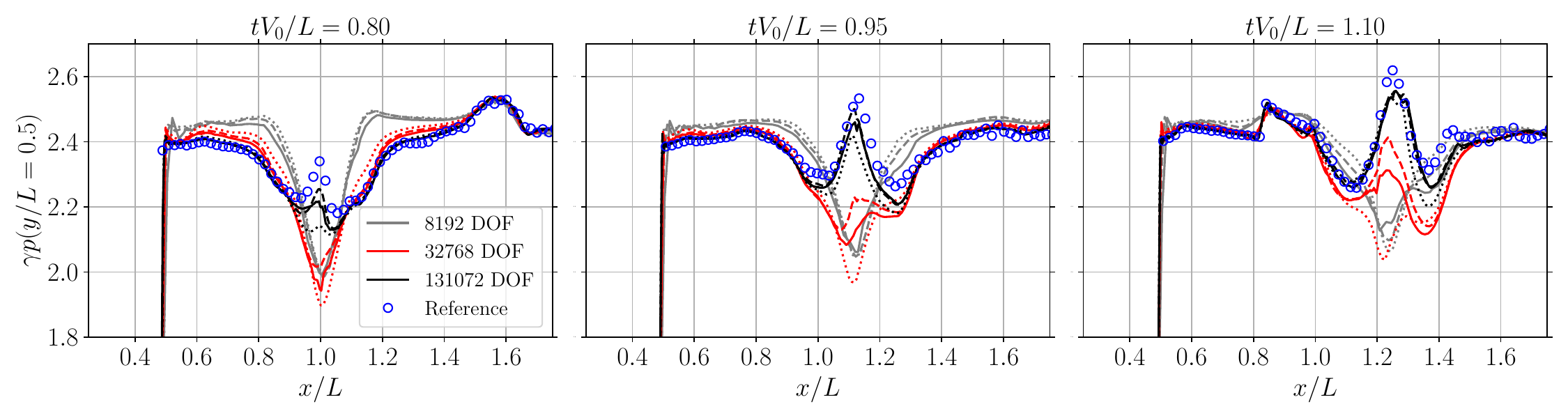}
\caption{\reviewerB{Pressure field at the center of the domain resulting from a strong shock-strong vortex interaction at various instants in time. Different grid resolution levels, measured by the total number of degrees-of-freedom (DOF), are compared against a reference solution (figure \ref{fig:ShockVortexPhys}). The different polynomial orders are distinguished by their particular line styles: low. $N=3$, is shown in dotted lines (\sampleline{dotted}), intermediate, $N=7$, in dashed lines (\sampleline{dashed}) and high, $N=15$, in solid lines (\sampleline{}).}
}\label{fig:ShockVortexCenterline} 
\end{figure}

\reviewerB{Moving on, a more quantitative examination of the resolution power for each level of total DOF and polynomial order is provided in figure \ref{fig:ShockVortexCenterline}. The pressure field at the center of the domain is extracted for each simulation at different time instants and compared against a LSV reference solution performed with approximately $2.1$ million DOF (figure \ref{fig:ShockVortexPhys}). Once more it can be observed that intermediate polynomial order of $N = 7$ is closer to the reference solution for a given total DOF level in comparison with the others, followed by $N=3$ and $N=15$. Additionally, it can be observed that results gathered from the simulations performed with approximately $130000$ DOF follow closely the ones from the converged reference solution, 16 times finer.}

\subsection{Double Mach Reflection}  \label{subsec:MachReflection}
The double Mach reflection problem of \citet{Woodward_JCP_1984} is a challenging test case for the 2D unsteady Euler equations. A Mach 10 shock is introduced in the computational domain $\Omega = [0,4]\times[0,1]$ at a $60$-degree angle with respect to the lower boundary. The left and right states are defined as 

\begin{minipage}[l]{.22\textwidth}
\begin{equation} \nonumber
\rho = 
\begin{cases}
\rho_l = \frac{(\gamma + 1)M^2}{2 + (\gamma - 1)M^2}, \\
\rho_r = 1,
\end{cases}
\end{equation}
\end{minipage}
\begin{minipage}{.22\textwidth}
\begin{equation} \nonumber
u = 
\begin{cases}
u_l = V_l \cos(30) ,\\
u_r =  0,
\end{cases}
\end{equation}
\end{minipage}
\begin{minipage}{.22\textwidth}
\begin{equation} \nonumber
v = 
\begin{cases}
v_l = -V_l \sin(30) ,\\
v_r =  0,
\end{cases}
\end{equation}
\end{minipage}
\begin{minipage}[r]{.22\textwidth}
\begin{equation}
p = 
\begin{cases}
p_l =  \frac{2 \gamma M^2 - (\gamma - 1)}{\gamma(\gamma + 1)},\\
p_r = 1/\gamma,
\end{cases}
\end{equation}
\end{minipage}

\noindent where $V_l = M - a_r \sqrt{\frac{(\gamma - 1) M^2 + 2}{2 \gamma M^2 - (\gamma - 1)}}$, $a_r  = \sqrt{\gamma \frac{p_{r}}{\rho_{r}}}$ is the speed of sound at the right state. In the current simulations the value $\gamma = 1.4$ was considered. At time $t=0$, the shock front touches the bottom boundary at $x = 1/6$ and is kept attached by assigning the left state values to the region where $x < 1/6$. For $x \in [1/6, 4]$, a reflecting wall boundary condition is applied at the bottom wall. At the top wall, a time dependent boundary condition that follows the shock front at $x_s = 1/6 + \frac{1 + 2Mt}{\sqrt{3}}$ is used to apply the correct left or right states. Ultimately, inflow and outflow conditions are assigned to the left and right boundaries, respectively. Such a setup was used by \citet{Kuzmin_JCP_2014} to study slope limiting in discontinuous Galerkin methods and by \citet{Asthana_JCP_2015} to analyze the Fourier-spectral filtering method in flux reconstruction settings.

Due to the substantial pressure jump imposed by the Mach 10 shock, in the order of 100, the positivity of the pressure field was difficult to maintain for a high-order discretization. \reviewerA{It is shown in \citet{Zhang_JCP_2017} that the use of SSP Runge-Kutta time advancement together with a suitable positivity preserving flux is able to ensure a \emph{weak positivity} property for the compressible Navier-Stokes equations. That is, given a sufficiently small enough time step, the positivity of the average of the cell values can be assured. Furthermore, if one analyzes the filtered compressible Navier-Stokes system of equations, one can observe that it has the same structure as the original equations with the addition of extra dissipative terms that come from the sub-filter scale (SFS) closure. Since the SFS terms have the same mathematical structure as the original stress tensor and heat flux vector, it is possible to construct a positivity-preserving flux that accounts for the SFS contributions. 

This, though, does not ensure pointwise positivity and, in \citet{Zhang_JCP_2017}, a scaling limiter that suppresses the solution without changing the average value is used to achieve the global positivity property. Although this procedure is, in theory, compatible with the current \modelAcronym~flux reconstruction-based implementation because it allows discontinuous solutions at cell interfaces, this step was not applied in the current work. Additionally, although it would be interesting to test the use of the developed positivity-preserving flux for the compressible Navier-Stokes equations \citep{Zhang_JCP_2017} in the current setup, the original choice of adopting the Lax-Friedrichs flux was retained and a study on the different numerical fluxes is left for future work. Ultimately, the increase in upwind degree that would be introduced by the novel flux formulation was substituted by a higher dissipation magnitude in this particular test case.}

To ensure numerical stability in the current double Mach reflection $4$-th order runs, it was necessary to increase the magnitude of the components of the dissipation tensor, $\mathcal{D}^{ij}$ \eqref{eqn:Dissipation}, by a factor of 3.  \citet{Asthana_JCP_2015} also reports the necessity of intensifying the added dissipation to simulate the double mach reflection stably, what was done by decreasing the threshold for the discontinuity sensor. A point for further investigation should be the effect of considering the Reynolds-based filtered equations, which would lead to different sub-filter flux terms that include a SFS density flux term and could possibly lead to implementations with higher numerical stability.

\begin{figure}[ht]
\centering
\includegraphics[width=.76\linewidth]{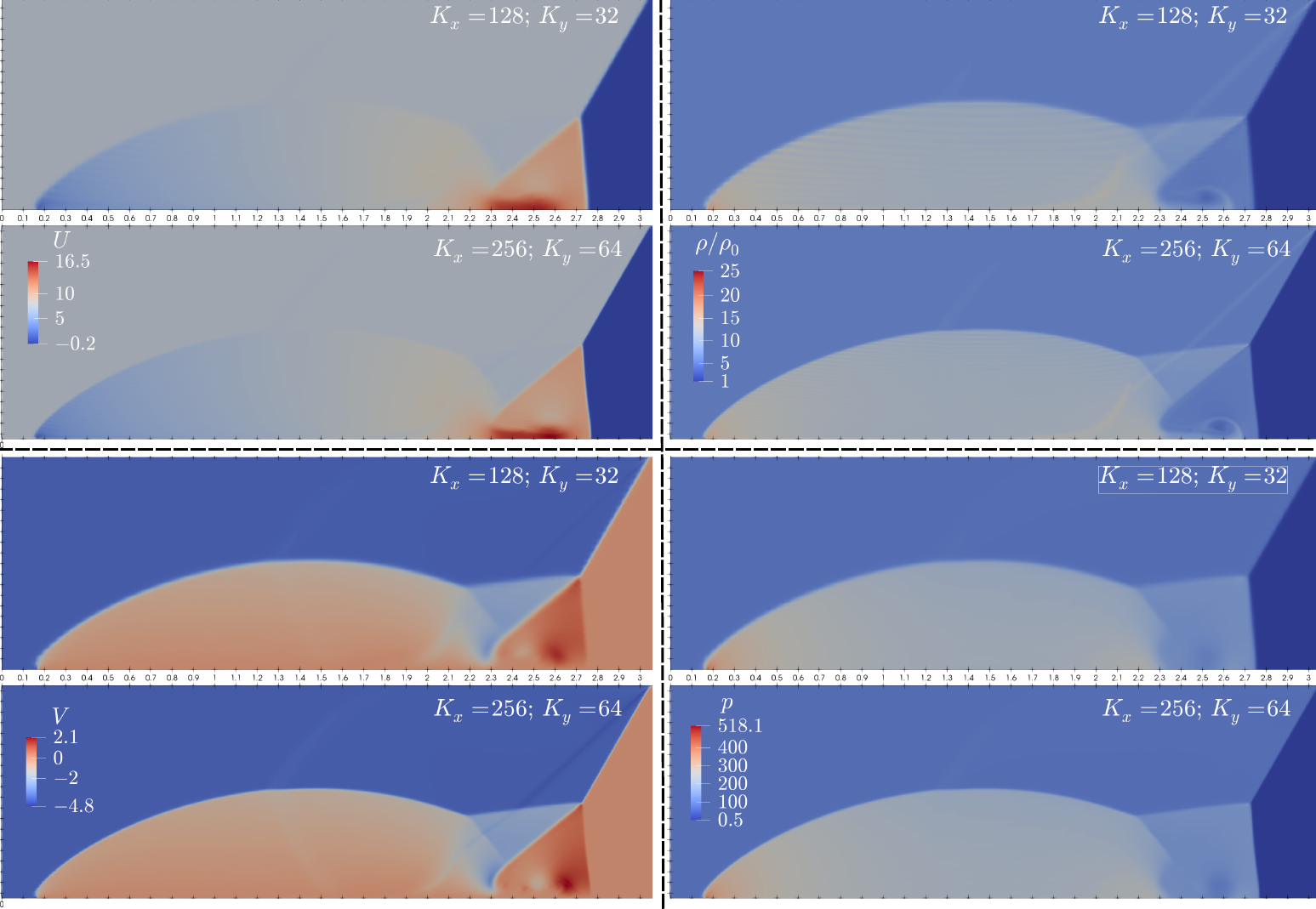}
\caption{Simulations of the reflection of a Mach 10 shock performed with $N=4$ polynomial order in both directions and $K_x \times  K_y= [128\times32,256\times64]$ cells. Results for the pressure, density and velocity fields are shown at time $t = 0.2$.
}\label{fig:mach10} 
\end{figure}

Figure \ref{fig:mach10} gathers the results for the pressure, density and velocity fields for the double Mach reflection problem at $t = 0.2$, when simulations are performed with a $4$-th order polynomial and $K_x \times  K_y= [128\times32,256\times64]$ cells. It can be observed that a high speed jet is formed at the center of the domain near the shock front and, in turn, the induced shear causes the formation of a vortex, most evidenced in the density field. Additionally, the grid refinement leads to sharper gradients at the shock boundaries, higher resolution of small scales in the vortex region, and decreased spurious density oscillations in the density field. The finest simulation here uses 409,600 degrees-of-freedom, about 2.5 times lower than \citet{Asthana_JCP_2015} for qualitatively similar results. Moreover, the finest simulation carried in the current work recovers considerably more flow dynamic details when compared to the finest resolution simulation performed by \citet{Kuzmin_JCP_2014}, which had 65,536 elements with 2nd order, equivalent to 589,824 DOFs.

\begin{figure}[ht]
\centering
\includegraphics[width=.5\linewidth]{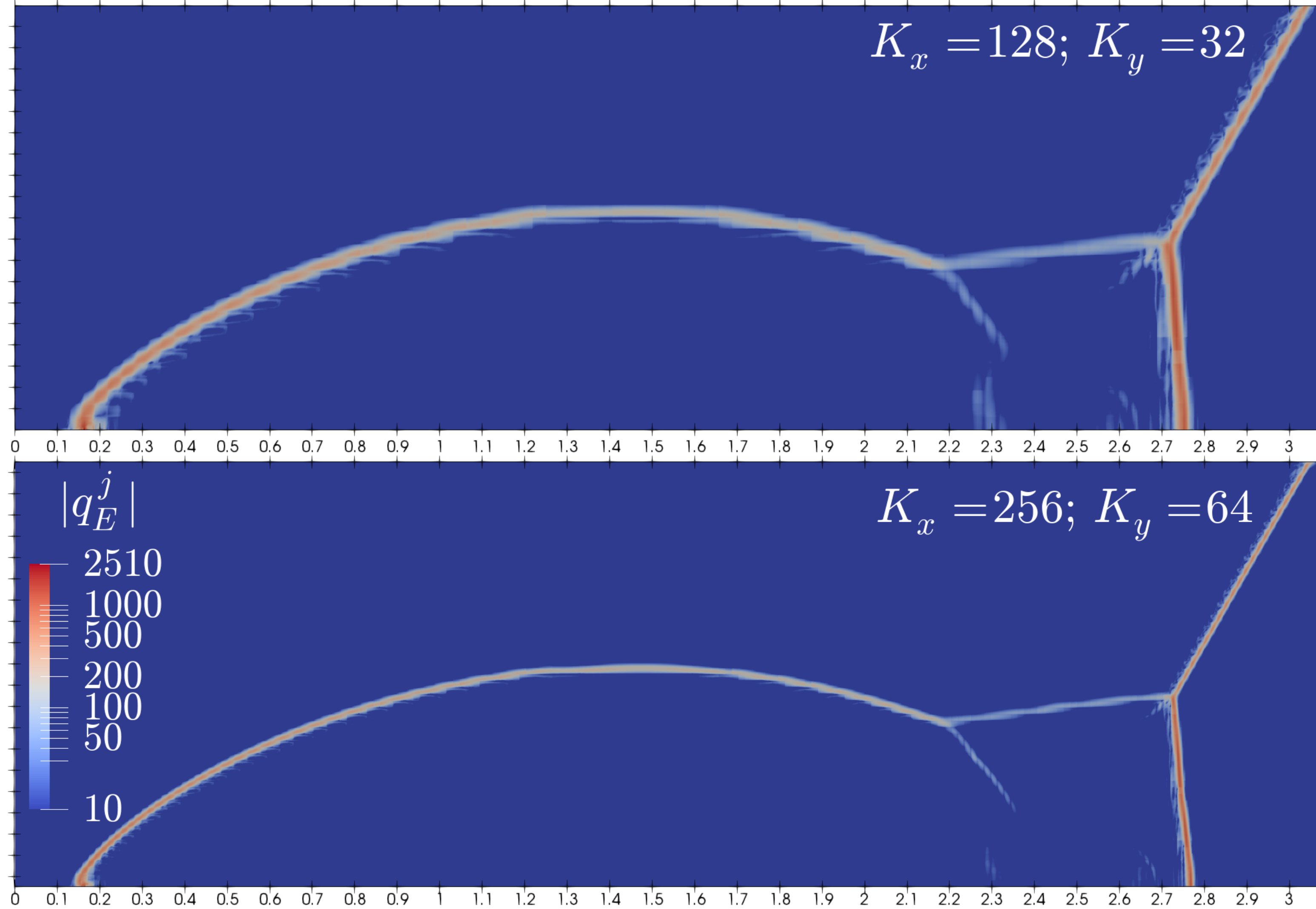}
\caption{Display of the magnitude of the sub filter scale total energy dissipation, $q^j_E =  \frac{1}{2} \left( \frac{\gamma \pi^j}{\gamma - 1} + \overline { \rho} C_p q^{j} \right)$, field.
}\label{fig:qMach10} 
\end{figure}

Ultimately, an illustration of the \modelAcronym~closure spatial arrangement is shown in figure \ref{fig:qMach10}. The sub filter scale total energy dissipation,

\be
q^j_E =  \frac{1}{2} \left( \frac{\gamma \pi^j}{\gamma - 1} + \overline { \rho} C_p q^{j} \right),
\ee

\noindent is highly concentrated near the discontinuity region since it is only active in elements where the nonlinear strength is big enough to reach the resolution limit and be detected by the energy at the cutoff operation \eqref{eqn:spectralKE}. The element localization is reenforced by the grid refinement step.

%-------------------------------------------------------------------------------------------------------------------------------------------------------------------------%

%-------------------------------------------------------------------------------------------------------------------------------------------------------------------------%
%!TEX root = ../SousaLSV_JCP_2022.tex

\section{Conclusion} \label{sec:Conclusion}

A novel technique for discontinuity capturing for high-order Flux Reconstruction (FR) schemes named the \modelFullName~model (\modelAcronym), was introduced. It relies on solving the large scales present in shock-dominated flows while dissipating the small scales. This is achieved by solving for filtered versions of the original nonlinear conservation equations and modeling the energy flux to sub-filter scales as a dissipative term. Ultimately, it is a generalization of the mathematical framework used in Large Eddy Simulations (LES) to shock-dominated problems.

The novelty of the \modelAcronym~method relies on exploiting the connection between FR implementation and the Legendre polynomials to project the solution within each cell onto the naturally occurring Legendre set of hierarchical basis functions. This step is performed to both estimate the magnitude of the needed dissipative term as well as modulate the dissipation behavior at different scales. In consequence, the \modelAcronym~method is able to introduce a spectrally concentrated at small scales dissipative term that is active only in cells where nonlinear dynamics are important.

The \modelAcronym~closure was successfully tested in one- and two-dimensional shock-dominated problems, being able to outperform consistently  results shown in \citet{Asthana_JCP_2015}, by requiring less degrees of freedom to achieve the same level of accuracy. Additionally, although qualitatively similar results are recovered when compared against previous 1D shock/entropy wave interaction results by \citet{HagaKawai_JCP_2019}, the current method is able to surpass the 4-th order limitation existent in the approach discussed in the aforementioned publication.

The results gathered here show the high performance of the \modelAcronym~closure model in discontinuity capturing in high-order flux reconstruction settings. In the future, it is proposed to extend the methodology to model turbulence and shocks concomitantly, in the example of \citet{Sousa_ArXiv_2021}, who showed that a similar setup for high-order finite difference solvers can be perform both tasks at the same time.

%-------------------------------------------------------------------------------------------------------------------------------------------------------------------------%

\begin{appendix}
\reviewerA{
%!TEX root = ../SousaLSV_JCP_2022.tex

\section{Effects of the spectral modulation operation on the solution of 1D shock-dominated problems} \label{sec:Appendix}

In this section, the Sod shock tube \citep{SOD_1978_JCP} and the shock-entropy wave interaction introduced by \citet{shu1988efficient} are simulated using the \modelAcronym~closure with and without the spectral modulation operation \eqref{eqn:Mod} to showcase the effects that such a procedure causes to the final solution.

The 1D Euler system of equations augmented by the models for the SFS flux terms, i.e. the inviscid version of the Navier-Stokes system of equations \eqref{eqn:dens} - \eqref{eqn:ene}, is solved via the flux reconstruction scheme discussed in the section \ref{sec:FRLSV} and advanced in time using a 3rd-order strong stability preserving (SSP) Runge-Kutta \citep{gottlieb2001strong} method. The relevant initial conditions for the Sod shock tube and the Shu-Osher problems can be found in sections \ref{subsec:Sod} and \ref{subsec:ShOsher}, respectively.

\begin{figure}[ht]
\centering
\includegraphics[width=.9\linewidth]{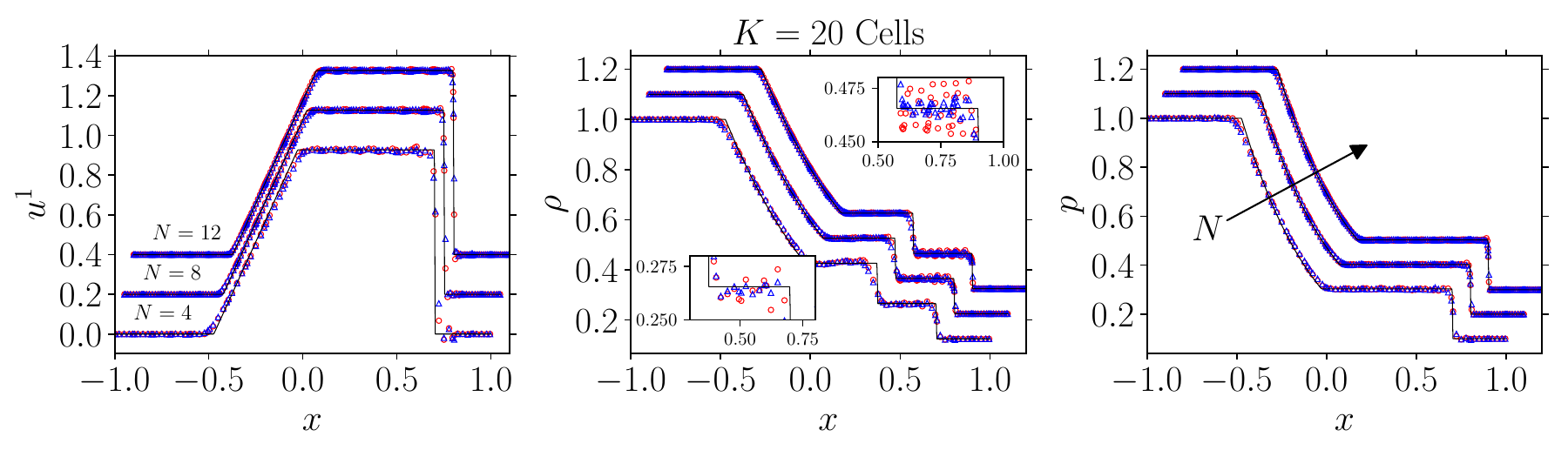}
\includegraphics[width=.9\linewidth]{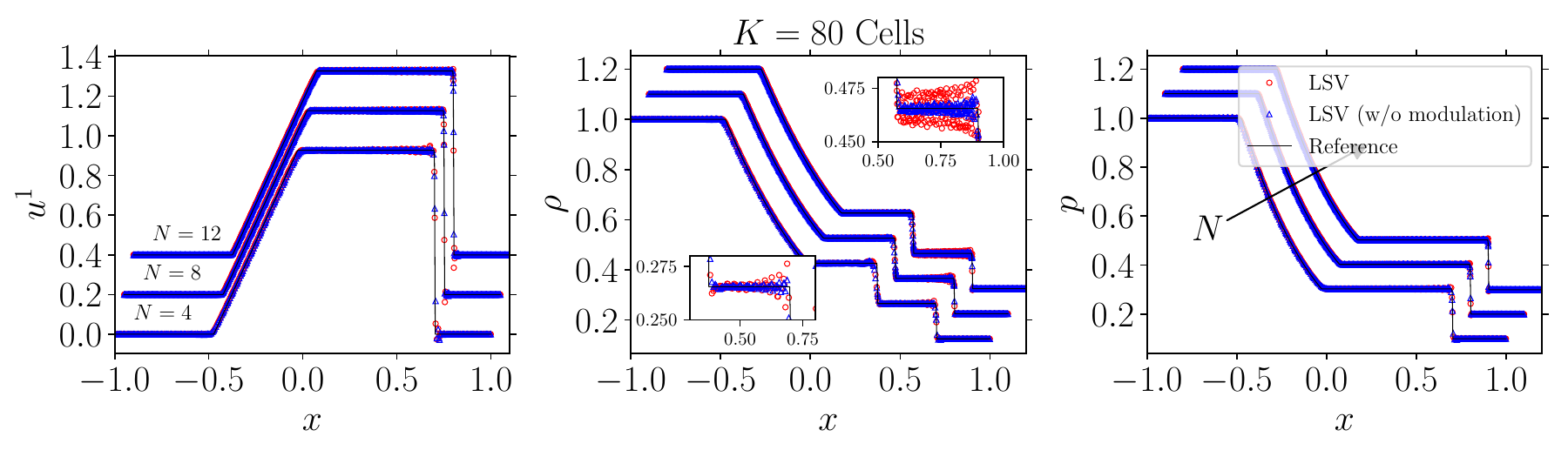}
\caption{\reviewerA{High-order flux reconstruction numerical solution of the Sod shock tube problem \citep{SOD_1978_JCP} using the \modelAcronym\ closure with and without the application of the spectral modulation procedure \eqref{eqn:Mod}. The domain is discretized with a combination of $K = [20,80]$ cells and $N=[4,8,12]$ polynomial orders and the initial condition is solved up to $t = 0.4$.}
}\label{fig:aposterioriSODnoMod} 
\end{figure}

Figure \ref{fig:aposterioriSODnoMod} shows that the increase in dissipation magnitude across all the resolved wavenumbers associated with skipping the spectral modulation operation (see figure \ref{fig:aprioriModulation}) is able to decrease the amplitude of the spurious oscillations present specially in the density field at the region between the main shock and the contact discontinuity. This decrease occurs for all combinations of number of cells and polynomial orders considered but its effectiveness is particularly observed at $K=80$ cells and $N = 12$ order. At such a resolution, the simulation without the modulation operation is capable of decreasing the observed high-frequency oscillations' magnitude by a factor of 3.

However, the increase in the added dissipation across the resolved spectrum can also lead to dissipation of physical small-scale oscillations such as the ones present in the Shu-Osher problem. Figure \ref{fig:aposterioriShuOsherNoMod} shows that the simulations performed without the spectral modulation operation are consistently further from the reference solution in comparison with the default \modelAcronym~closure setup for all grid arrangements considered. For example, at the $K = 40$ and $N = 8$ resolution, the full \modelAcronym~results for the acoustic waves trailing the main shock are fairly close to the grid resolved simulation while the ones achieved by skipping the spectral modulation operation display significantly lower magnitude.

Ultimately, the choice was made to accurately retain as much spectral information as possible in the resolved solution for a given grid resolution in the \modelAcronym~ methodology, even though such choice leads the occurrence of low-amplitude high-frequency oscillations in the solution.

\begin{figure}[ht]
\centering
\includegraphics[width=.3\linewidth]{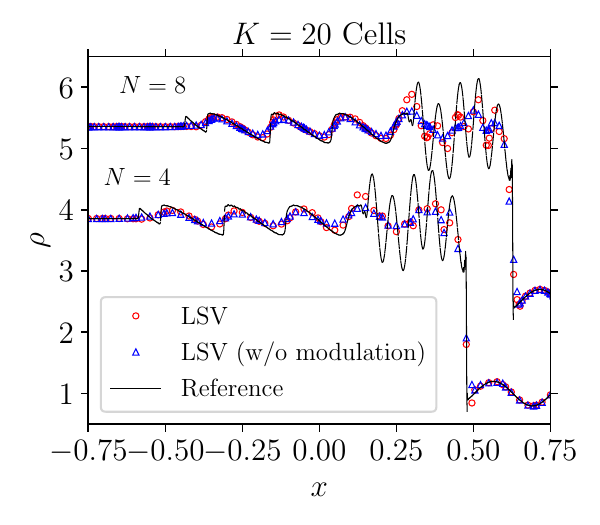}
\includegraphics[width=.3\linewidth]{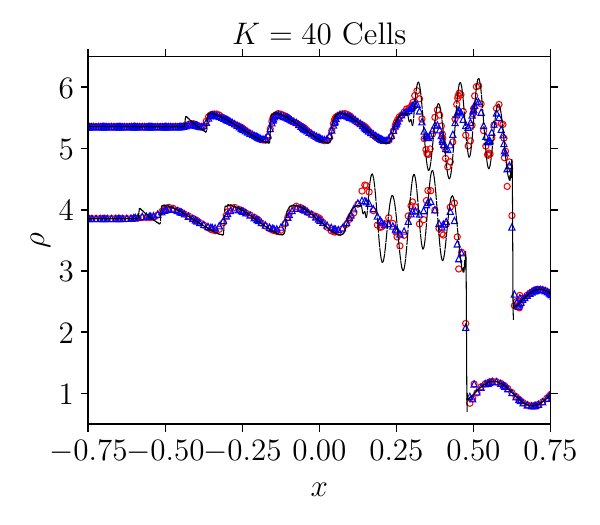}
\includegraphics[width=.3\linewidth]{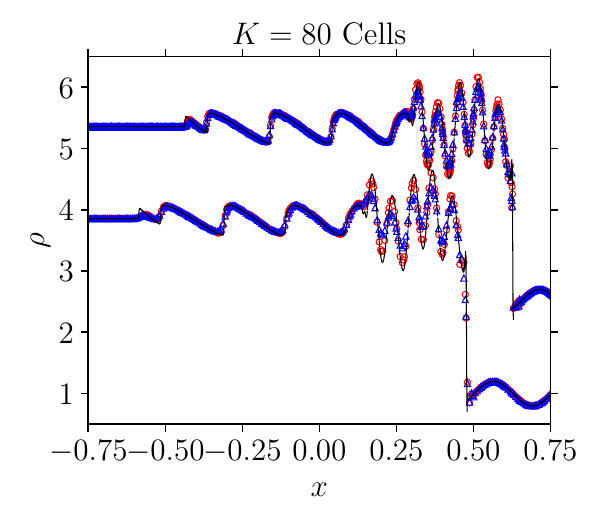}
\caption{\reviewerA{High-order flux reconstruction numerical solution of the Shu-Osher shock-entropy wave interaction  \citep{shu1988efficient} using the \modelAcronym\ closure with and without the application of the spectral modulation procedure \eqref{eqn:Mod}. The domain is discretized with a combination of $K = [20,40,80]$ cells and $N=[4,8]$ polynomial orders and the initial condition is solved up to $t = 0.36$. Numerical results at the solution points are compared against a DNS performed with 500 cells and order 3, shown in fine solid black lines.}
}\label{fig:aposterioriShuOsherNoMod} 
\end{figure}

}
\end{appendix}

\section*{Acknowledgments}
Victor Sousa and Carlo Scalo acknowledge the computational support of the Rosen Center for Advanced Computing (RCAC) at Purdue and of the U.S. Air Force Research Laboratory (AFRL) DoD Supercomputing Resource Center (DSRC), via allocation under the subproject AFOSR43032009. This project was funded by the Air Force Office of Scientific Research (AFOSR) Young Investigator Award (YIP) Grant (FA9550-18-271-0292), the Office of Naval Research YIP (N000142012662) as well as the ONR Grant No. N00014-21-1-2475. Victor Sousa also acknowledges the support of the Lynn Fellowship administered by the interdisciplinary Computational Science and Engineering (CS\&E) graduate program at Purdue University. 

\bibliography{./references.bib}

\newcommand{\noop}[1]{}
\begin{thebibliography}{60}
\expandafter\ifx\csname natexlab\endcsname\relax\def\natexlab#1{#1}\fi
\providecommand{\url}[1]{\texttt{#1}}
\providecommand{\href}[2]{#2}
\providecommand{\path}[1]{#1}
\providecommand{\DOIprefix}{doi:}
\providecommand{\ArXivprefix}{arXiv:}
\providecommand{\URLprefix}{URL: }
\providecommand{\Pubmedprefix}{pmid:}
\providecommand{\doi}[1]{\href{http://dx.doi.org/#1}{\path{#1}}}
\providecommand{\Pubmed}[1]{\href{pmid:#1}{\path{#1}}}
\providecommand{\bibinfo}[2]{#2}
\ifx\xfnm\relax \def\xfnm[#1]{\unskip,\space#1}\fi
%Type = Article
\bibitem[{Asthana et~al.(2015)Asthana, L{\'o}pez-Morales and
  Jameson}]{Asthana_JCP_2015}
\bibinfo{author}{Asthana, K.}, \bibinfo{author}{L{\'o}pez-Morales, M.R.},
  \bibinfo{author}{Jameson, A.}, \bibinfo{year}{2015}.
\newblock \bibinfo{title}{Non-linear stabilization of high-order flux
  reconstruction schemes via {F}ourier-spectral filtering}.
\newblock \bibinfo{journal}{Journal of Computational Physics}
  \bibinfo{volume}{303}, \bibinfo{pages}{269--294}.
%Type = Article
\bibitem[{Barter and Darmofal(2010)}]{Barter_JCP_2010}
\bibinfo{author}{Barter, G.E.}, \bibinfo{author}{Darmofal, D.L.},
  \bibinfo{year}{2010}.
\newblock \bibinfo{title}{{Shock capturing with PDE-based artificial viscosity
  for DGFEM: Part I. Formulation}}.
\newblock \bibinfo{journal}{Journal of Computational Physics}
  \bibinfo{volume}{229}, \bibinfo{pages}{1810--1827}.
%Type = Article
\bibitem[{Bassi and Rebay(1997)}]{bassi1997BR1high}
\bibinfo{author}{Bassi, F.}, \bibinfo{author}{Rebay, S.}, \bibinfo{year}{1997}.
\newblock \bibinfo{title}{A high-order accurate discontinuous finite element
  method for the numerical solution of the compressible {N}avier-{S}tokes
  equations}.
\newblock \bibinfo{journal}{J. Comput. Phys.} \bibinfo{volume}{131},
  \bibinfo{pages}{267--279}.
%Type = Article
\bibitem[{Batten et~al.(1997)Batten, Clarke, Lambert and
  Causon}]{Batten_SIAM_1997}
\bibinfo{author}{Batten, P.}, \bibinfo{author}{Clarke, N.},
  \bibinfo{author}{Lambert, C.}, \bibinfo{author}{Causon, D.M.},
  \bibinfo{year}{1997}.
\newblock \bibinfo{title}{{On the choice of wavespeeds for the HLLC Riemann
  solver}}.
\newblock \bibinfo{journal}{SIAM Journal on Scientific Computing}
  \bibinfo{volume}{18}, \bibinfo{pages}{1553--1570}.
%Type = Article
\bibitem[{Biswas et~al.(1994)Biswas, Devine and Flaherty}]{Biswas_ANM_1994}
\bibinfo{author}{Biswas, R.}, \bibinfo{author}{Devine, K.D.},
  \bibinfo{author}{Flaherty, J.E.}, \bibinfo{year}{1994}.
\newblock \bibinfo{title}{Parallel, adaptive finite element methods for
  conservation laws}.
\newblock \bibinfo{journal}{Applied Numerical Mathematics}
  \bibinfo{volume}{14}, \bibinfo{pages}{255--283}.
%Type = Article
\bibitem[{Burbeau et~al.(2001)Burbeau, Sagaut and Bruneau}]{Burbeau_JCP_2001}
\bibinfo{author}{Burbeau, A.}, \bibinfo{author}{Sagaut, P.},
  \bibinfo{author}{Bruneau, C.H.}, \bibinfo{year}{2001}.
\newblock \bibinfo{title}{{A problem-independent limiter for high-order
  Runge--Kutta discontinuous Galerkin methods}}.
\newblock \bibinfo{journal}{Journal of Computational Physics}
  \bibinfo{volume}{169}, \bibinfo{pages}{111--150}.
%Type = Article
\bibitem[{Chollet and Lesieur(1981)}]{chollet1981parameterization}
\bibinfo{author}{Chollet, J.}, \bibinfo{author}{Lesieur, M.},
  \bibinfo{year}{1981}.
\newblock \bibinfo{title}{Parameterization of small scales of three-dimensional
  isotropic turbulence utilizing spectral closures}.
\newblock \bibinfo{journal}{J. Atmo. Sci.} \bibinfo{volume}{38},
  \bibinfo{pages}{2747--2757}.
%Type = Article
\bibitem[{Clark et~al.(1979)Clark, Ferziger and Reynolds}]{Clark_JFM_1979}
\bibinfo{author}{Clark, R.A.}, \bibinfo{author}{Ferziger, J.H.},
  \bibinfo{author}{Reynolds, W.C.}, \bibinfo{year}{1979}.
\newblock \bibinfo{title}{Evaluation of subgrid-scale models using an
  accurately simulated turbulent flow}.
\newblock \bibinfo{journal}{Journal of fluid mechanics} \bibinfo{volume}{91},
  \bibinfo{pages}{1--16}.
%Type = Article
\bibitem[{Cockburn and Shu(1989)}]{Cockburn_MC_1989}
\bibinfo{author}{Cockburn, B.}, \bibinfo{author}{Shu, C.W.},
  \bibinfo{year}{1989}.
\newblock \bibinfo{title}{{TVB} {Runge}-{Kutta} local projection discontinuous
  {Galerkin} finite element method for conservation laws. {II}. {General}
  framework}.
\newblock \bibinfo{journal}{Math. Comp.} \bibinfo{volume}{52},
  \bibinfo{pages}{411--411}.
\newblock \URLprefix
  \url{http://www.ams.org/jourcgi/jour-getitem?pii=S0025-5718-1989-0983311-4},
  \DOIprefix\doi{10.1090/S0025-5718-1989-0983311-4}.
%Type = Article
\bibitem[{Cook(2007)}]{Cook_PoF_2007}
\bibinfo{author}{Cook, A.W.}, \bibinfo{year}{2007}.
\newblock \bibinfo{title}{{Artificial Fluid Properties for Large-Eddy
  Simulation of Compressible Turbulent Mixing}}.
\newblock \bibinfo{journal}{Physics of fluids} \bibinfo{volume}{19},
  \bibinfo{pages}{055103}.
%Type = Article
\bibitem[{Dosanjh and Weeks(1965)}]{dosanjh_ShockVortex_1965}
\bibinfo{author}{Dosanjh, D.S.}, \bibinfo{author}{Weeks, T.M.},
  \bibinfo{year}{1965}.
\newblock \bibinfo{title}{Interaction of a starting vortex as well as a vortex
  street with a traveling shock wave}.
\newblock \bibinfo{journal}{AIAA journal} \bibinfo{volume}{3},
  \bibinfo{pages}{216--223}.
%Type = Article
\bibitem[{Ellzey et~al.(1995)Ellzey, Henneke, Picone and
  Oran}]{Ellzey_PoF_1995}
\bibinfo{author}{Ellzey, J.L.}, \bibinfo{author}{Henneke, M.R.},
  \bibinfo{author}{Picone, J.M.}, \bibinfo{author}{Oran, E.S.},
  \bibinfo{year}{1995}.
\newblock \bibinfo{title}{The interaction of a shock with a vortex: shock
  distortion and the production of acoustic waves}.
\newblock \bibinfo{journal}{Physics of Fluids} \bibinfo{volume}{7},
  \bibinfo{pages}{172--184}.
%Type = Article
\bibitem[{Germano et~al.(1991)Germano, Piomelli, Moin and Cabot}]{GermanoPMC91}
\bibinfo{author}{Germano, M.}, \bibinfo{author}{Piomelli, U.},
  \bibinfo{author}{Moin, P.}, \bibinfo{author}{Cabot, W.},
  \bibinfo{year}{1991}.
\newblock \bibinfo{title}{{A dynamic subgrid-scale eddy viscosity model}}.
\newblock \bibinfo{journal}{Phys.\ Fluids A} \bibinfo{volume}{3},
  \bibinfo{pages}{1760--1765}.
%Type = Article
\bibitem[{Godunov and Bohachevsky(1959)}]{Godunov_1959}
\bibinfo{author}{Godunov, S.}, \bibinfo{author}{Bohachevsky, I.},
  \bibinfo{year}{1959}.
\newblock \bibinfo{title}{Finite difference method for numerical computation of
  discontinuous solutions of the equations of fluid dynamics}.
\newblock \bibinfo{journal}{Matemati{\v{c}}eskij sbornik} \bibinfo{volume}{47},
  \bibinfo{pages}{271--306}.
%Type = Article
\bibitem[{Gottlieb and Shu(1997)}]{Gottlieb_SIAM_1997}
\bibinfo{author}{Gottlieb, D.}, \bibinfo{author}{Shu, C.W.},
  \bibinfo{year}{1997}.
\newblock \bibinfo{title}{On the {G}ibbs phenomenon and its resolution}.
\newblock \bibinfo{journal}{SIAM review} \bibinfo{volume}{39},
  \bibinfo{pages}{644--668}.
%Type = Article
\bibitem[{Gottlieb et~al.(2001)Gottlieb, Shu and Tadmor}]{gottlieb2001strong}
\bibinfo{author}{Gottlieb, S.}, \bibinfo{author}{Shu, C.},
  \bibinfo{author}{Tadmor, E.}, \bibinfo{year}{2001}.
\newblock \bibinfo{title}{Strong stability-preserving high-order time
  discretization methods}.
\newblock \bibinfo{journal}{SIAMR} \bibinfo{volume}{43},
  \bibinfo{pages}{89--112}.
%Type = Article
\bibitem[{Grasso and Pirozzoli(2000)}]{Grasso_ShockVortex_2000}
\bibinfo{author}{Grasso, F.}, \bibinfo{author}{Pirozzoli, S.},
  \bibinfo{year}{2000}.
\newblock \bibinfo{title}{Shock-wave--vortex interactions: shock and vortex
  deformations, and sound production}.
\newblock \bibinfo{journal}{Theoretical and Computational Fluid Dynamics}
  \bibinfo{volume}{13}, \bibinfo{pages}{421--456}.
%Type = Article
\bibitem[{Haga and Kawai(2019)}]{HagaKawai_JCP_2019}
\bibinfo{author}{Haga, T.}, \bibinfo{author}{Kawai, S.}, \bibinfo{year}{2019}.
\newblock \bibinfo{title}{On a robust and accurate localized artificial
  diffusivity scheme for the high-order flux-reconstruction method}.
\newblock \bibinfo{journal}{Journal of Computational Physics}
  \bibinfo{volume}{376}, \bibinfo{pages}{534--563}.
%Type = Article
\bibitem[{Harten et~al.(1983)Harten, Lax and Leer}]{HLL_1983}
\bibinfo{author}{Harten, A.}, \bibinfo{author}{Lax, P.D.},
  \bibinfo{author}{Leer, B.v.}, \bibinfo{year}{1983}.
\newblock \bibinfo{title}{On upstream differencing and godunov-type schemes for
  hyperbolic conservation laws}.
\newblock \bibinfo{journal}{SIAM review} \bibinfo{volume}{25},
  \bibinfo{pages}{35--61}.
%Type = Article
\bibitem[{Huynh(2007)}]{huynh2007flux}
\bibinfo{author}{Huynh, H.}, \bibinfo{year}{2007}.
\newblock \bibinfo{title}{A {F}lux {R}econstruction approach to high-order
  schemes including discontinuous {G}alerkin methods}.
\newblock \bibinfo{journal}{AIAA Paper} \bibinfo{volume}{2007-4079},
  \bibinfo{pages}{1--42}.
\newblock \bibinfo{note}{18th AIAA Computational Fluid Dynamics Conference,
  Miami, FL, Jun.~25--28, 2007}.
%Type = Article
\bibitem[{Jameson et~al.(2012)Jameson, Vincent and
  Castonguay}]{Jameson_JSC_2012}
\bibinfo{author}{Jameson, A.}, \bibinfo{author}{Vincent, P.E.},
  \bibinfo{author}{Castonguay, P.}, \bibinfo{year}{2012}.
\newblock \bibinfo{title}{On the non-linear stability of flux reconstruction
  schemes}.
\newblock \bibinfo{journal}{Journal of Scientific Computing}
  \bibinfo{volume}{50}, \bibinfo{pages}{434--445}.
%Type = Article
\bibitem[{Jordan(1999)}]{Jordan_JCP_1999}
\bibinfo{author}{Jordan, S.A.}, \bibinfo{year}{1999}.
\newblock \bibinfo{title}{A large-eddy simulation methodology in generalized
  curvilinear coordinates}.
\newblock \bibinfo{journal}{Journal of Computational Physics}
  \bibinfo{volume}{148}, \bibinfo{pages}{322--340}.
%Type = Article
\bibitem[{Karamanos and Karniadakis(2000)}]{Karamanos_JCP_2000}
\bibinfo{author}{Karamanos, G.}, \bibinfo{author}{Karniadakis, G.E.},
  \bibinfo{year}{2000}.
\newblock \bibinfo{title}{A spectral vanishing viscosity method for large-eddy
  simulations}.
\newblock \bibinfo{journal}{Journal of Computational Physics}
  \bibinfo{volume}{163}, \bibinfo{pages}{22--50}.
%Type = Book
\bibitem[{Karniadakis and Sherwin(2013)}]{Karniadakis2013spectral}
\bibinfo{author}{Karniadakis, G.}, \bibinfo{author}{Sherwin, S.},
  \bibinfo{year}{2013}.
\newblock \bibinfo{title}{Spectral/hp element methods for computational fluid
  dynamics}.
\newblock \bibinfo{publisher}{Oxford University Press}.
%Type = Article
\bibitem[{Kawai and Lele(2008)}]{Kawai_JCP_2008}
\bibinfo{author}{Kawai, S.}, \bibinfo{author}{Lele, S.K.},
  \bibinfo{year}{2008}.
\newblock \bibinfo{title}{{Localized Artificial Diffusivity Scheme for
  Discontinuity Capturing on Curvilinear Meshes}}.
\newblock \bibinfo{journal}{Journal of Computational Physics}
  \bibinfo{volume}{227}, \bibinfo{pages}{9498--9526}.
%Type = Article
\bibitem[{Kawai et~al.(2010)Kawai, Shankar and Lele}]{Kawai_JCP_2010}
\bibinfo{author}{Kawai, S.}, \bibinfo{author}{Shankar, S.K.},
  \bibinfo{author}{Lele, S.K.}, \bibinfo{year}{2010}.
\newblock \bibinfo{title}{Assessment of localized artificial diffusivity scheme
  for large-eddy simulation of compressible turbulent flows}.
\newblock \bibinfo{journal}{Journal of Computational Physics}
  \bibinfo{volume}{229}, \bibinfo{pages}{1739--1762}.
%Type = Article
\bibitem[{Kirby and Karniadakis(2002)}]{Kirby_JFE_2002}
\bibinfo{author}{Kirby, R.M.}, \bibinfo{author}{Karniadakis, G.E.},
  \bibinfo{year}{2002}.
\newblock \bibinfo{title}{Coarse resolution turbulence simulations with
  spectral vanishing viscosity—large-eddy simulations ({SVV-LES})}.
\newblock \bibinfo{journal}{J. Fluids Eng.} \bibinfo{volume}{124},
  \bibinfo{pages}{886--891}.
%Type = Article
\bibitem[{Kopriva and Kolias(1996)}]{kopriva1996conservative}
\bibinfo{author}{Kopriva, D.A.}, \bibinfo{author}{Kolias, J.H.},
  \bibinfo{year}{1996}.
\newblock \bibinfo{title}{A conservative staggered-grid {C}hebyshev multidomain
  method for compressible flows}.
\newblock \bibinfo{journal}{J. Comput. Phys.} \bibinfo{volume}{125},
  \bibinfo{pages}{244--261}.
%Type = Article
\bibitem[{Kraichnan(1976)}]{kraichnan1976eddy}
\bibinfo{author}{Kraichnan, R.}, \bibinfo{year}{1976}.
\newblock \bibinfo{title}{Eddy viscosity in two and three dimensions}.
\newblock \bibinfo{journal}{J. Atmo. Sci.} \bibinfo{volume}{33},
  \bibinfo{pages}{1521--1536}.
%Type = Article
\bibitem[{Krivodonova(2007)}]{Krivodonova_JCP_2007}
\bibinfo{author}{Krivodonova, L.}, \bibinfo{year}{2007}.
\newblock \bibinfo{title}{Limiters for high-order discontinuous {Galerkin}
  methods}.
\newblock \bibinfo{journal}{Journal of Computational Physics}
  \bibinfo{volume}{226}, \bibinfo{pages}{879--896}.
%Type = Article
\bibitem[{Kuzmin(2014)}]{Kuzmin_JCP_2014}
\bibinfo{author}{Kuzmin, D.}, \bibinfo{year}{2014}.
\newblock \bibinfo{title}{Hierarchical slope limiting in explicit and implicit
  discontinuous {Galerkin} methods}.
\newblock \bibinfo{journal}{Journal of Computational Physics}
  \bibinfo{volume}{257}, \bibinfo{pages}{1140--1162}.
%Type = Article
\bibitem[{Luo et~al.(2007)Luo, Baum and L{\"o}hner}]{Luo_JCP_2007}
\bibinfo{author}{Luo, H.}, \bibinfo{author}{Baum, J.D.},
  \bibinfo{author}{L{\"o}hner, R.}, \bibinfo{year}{2007}.
\newblock \bibinfo{title}{{A Hermite WENO-based limiter for discontinuous
  Galerkin method on unstructured grids}}.
\newblock \bibinfo{journal}{Journal of Computational Physics}
  \bibinfo{volume}{225}, \bibinfo{pages}{686--713}.
%Type = Article
\bibitem[{Maday et~al.(1993)Maday, Kaber and Tadmor}]{Maday_JNA_1993}
\bibinfo{author}{Maday, Y.}, \bibinfo{author}{Kaber, S.M.O.},
  \bibinfo{author}{Tadmor, E.}, \bibinfo{year}{1993}.
\newblock \bibinfo{title}{Legendre pseudospectral viscosity method for
  nonlinear conservation laws}.
\newblock \bibinfo{journal}{SIAM Journal on Numerical Analysis}
  \bibinfo{volume}{30}, \bibinfo{pages}{321--342}.
%Type = Article
\bibitem[{Moin et~al.(1991)Moin, Squires, Cabot and Lee}]{moin1991dynamic}
\bibinfo{author}{Moin, P.}, \bibinfo{author}{Squires, K.},
  \bibinfo{author}{Cabot, W.}, \bibinfo{author}{Lee, S.}, \bibinfo{year}{1991}.
\newblock \bibinfo{title}{{A dynamic subgrid-scale model for compressible
  turbulence and scalar transport}}.
\newblock \bibinfo{journal}{Physics of Fluids A: Fluid Dynamics (1989-1993)}
  \bibinfo{volume}{3}, \bibinfo{pages}{2746--2757}.
%Type = Article
\bibitem[{Nagarajan et~al.(2007)Nagarajan, Lele and
  Ferziger}]{nagarajan2007leading}
\bibinfo{author}{Nagarajan, S.}, \bibinfo{author}{Lele, S.},
  \bibinfo{author}{Ferziger, J.}, \bibinfo{year}{2007}.
\newblock \bibinfo{title}{{Leading-Edge Effects in Bypass Transition}}.
\newblock \bibinfo{journal}{J.~Fluid~Mech.} \bibinfo{volume}{572},
  \bibinfo{pages}{471--504}.
%Type = Article
\bibitem[{Nagarajan et~al.(2003)Nagarajan, Lele and
  Ferziger}]{NagarajanLF_JCP_2003}
\bibinfo{author}{Nagarajan, S.}, \bibinfo{author}{Lele, S.K.},
  \bibinfo{author}{Ferziger, J.H.}, \bibinfo{year}{2003}.
\newblock \bibinfo{title}{{A Robust High-Order Compact Method for Large-Eddy
  Simulation}}.
\newblock \bibinfo{journal}{Journal of Computational Physics}
  \bibinfo{volume}{191}, \bibinfo{pages}{392--419}.
%Type = Inproceedings
\bibitem[{Nguyen et~al.(2007)Nguyen, Persson and Peraire}]{Nguyen_AIAA_2007}
\bibinfo{author}{Nguyen, N.}, \bibinfo{author}{Persson, P.O.},
  \bibinfo{author}{Peraire, J.}, \bibinfo{year}{2007}.
\newblock \bibinfo{title}{{RANS solutions using high order discontinuous
  Galerkin methods}}, in: \bibinfo{booktitle}{45th AIAA Aerospace Sciences
  Meeting and Exhibit}, p. \bibinfo{pages}{914}.
%Type = Article
\bibitem[{Normand and Lesieur(1992)}]{Normand_TCFD_1992}
\bibinfo{author}{Normand, X.}, \bibinfo{author}{Lesieur, M.},
  \bibinfo{year}{1992}.
\newblock \bibinfo{title}{Direct and large-eddy simulations of transition in
  the compressible boundary layer}.
\newblock \bibinfo{journal}{Theoretical and Computational Fluid Dynamics}
  \bibinfo{volume}{3}, \bibinfo{pages}{231--252}.
%Type = Article
\bibitem[{Persson and Peraire(2006)}]{persson2006sub}
\bibinfo{author}{Persson, P.O.}, \bibinfo{author}{Peraire, J.},
  \bibinfo{year}{2006}.
\newblock \bibinfo{title}{Sub-cell shock capturing for discontinuous {G}alerkin
  methods}.
\newblock \bibinfo{journal}{AIAA paper} \bibinfo{volume}{112},
  \bibinfo{pages}{2006}.
%Type = Article
\bibitem[{Piomelli et~al.(1988)Piomelli, Moin and
  Ferziger}]{PiomelliMF_PoF_1988}
\bibinfo{author}{Piomelli, U.}, \bibinfo{author}{Moin, P.},
  \bibinfo{author}{Ferziger, J.H.}, \bibinfo{year}{1988}.
\newblock \bibinfo{title}{{Model consistency in large eddy simulation of
  turbulent channel flows}}.
\newblock \bibinfo{journal}{Phys. Fluids} \bibinfo{volume}{31},
  \bibinfo{pages}{1884--1891}.
\newblock \DOIprefix\doi{10.1063/1.866635}.
%Type = Article
\bibitem[{Premasuthan et~al.(2014)Premasuthan, Liang and
  Jameson}]{Premasuthan_2014_I}
\bibinfo{author}{Premasuthan, S.}, \bibinfo{author}{Liang, C.},
  \bibinfo{author}{Jameson, A.}, \bibinfo{year}{2014}.
\newblock \bibinfo{title}{Computation of flows with shocks using the spectral
  difference method with artificial viscosity, {I}: basic formulation and
  application}.
\newblock \bibinfo{journal}{Computers \& Fluids} \bibinfo{volume}{98},
  \bibinfo{pages}{111--121}.
%Type = Article
\bibitem[{Qiu and Shu(2005)}]{Qiu_SIAM_2005}
\bibinfo{author}{Qiu, J.}, \bibinfo{author}{Shu, C.W.}, \bibinfo{year}{2005}.
\newblock \bibinfo{title}{Runge-kutta discontinuous {Galerkin} method using
  {WENO} limiters}.
\newblock \bibinfo{journal}{SIAM Journal on Scientific Computing}
  \bibinfo{volume}{26}, \bibinfo{pages}{907--929}.
%Type = Article
\bibitem[{Rault et~al.(2003)Rault, Chiavassa and Donat}]{Rault_JSC_2003}
\bibinfo{author}{Rault, A.}, \bibinfo{author}{Chiavassa, G.},
  \bibinfo{author}{Donat, R.}, \bibinfo{year}{2003}.
\newblock \bibinfo{title}{Shock-vortex interactions at high mach numbers}.
\newblock \bibinfo{journal}{Journal of Scientific Computing}
  \bibinfo{volume}{19}, \bibinfo{pages}{347--371}.
%Type = Book
\bibitem[{Shen et~al.(2011)Shen, Tang and Wang}]{Shen_SpectralBook_2011}
\bibinfo{author}{Shen, J.}, \bibinfo{author}{Tang, T.}, \bibinfo{author}{Wang,
  L.L.}, \bibinfo{year}{2011}.
\newblock \bibinfo{title}{Spectral methods: algorithms, analysis and
  applications}. volume~\bibinfo{volume}{41}.
\newblock \bibinfo{publisher}{Springer Science \& Business Media}.
%Type = Article
\bibitem[{Shu and Osher(1988)}]{shu1988efficient}
\bibinfo{author}{Shu, C.}, \bibinfo{author}{Osher, S.}, \bibinfo{year}{1988}.
\newblock \bibinfo{title}{Efficient implementation of essentially
  non-oscillatory shock-capturing schemes}.
\newblock \bibinfo{journal}{J. Comput. Phys.} \bibinfo{volume}{77},
  \bibinfo{pages}{439--471}.
%Type = Article
\bibitem[{Shu(1987)}]{Shu_MoC_1987}
\bibinfo{author}{Shu, C.W.}, \bibinfo{year}{1987}.
\newblock \bibinfo{title}{{TVB} uniformly high-order schemes for conservation
  laws}.
\newblock \bibinfo{journal}{Mathematics of Computation} \bibinfo{volume}{49},
  \bibinfo{pages}{105--121}.
%Type = Article
\bibitem[{Sidharth and Candler(2018)}]{Sidharth_JFM_2018}
\bibinfo{author}{Sidharth, G.}, \bibinfo{author}{Candler, G.V.},
  \bibinfo{year}{2018}.
\newblock \bibinfo{title}{Subgrid-scale effects in compressible
  variable-density decaying turbulence}.
\newblock \bibinfo{journal}{Journal of Fluid Mechanics} \bibinfo{volume}{846},
  \bibinfo{pages}{428--459}.
%Type = Article
\bibitem[{Sod(1978)}]{SOD_1978_JCP}
\bibinfo{author}{Sod, G.A.}, \bibinfo{year}{1978}.
\newblock \bibinfo{title}{A survey of several finite difference methods for
  systems of nonlinear hyperbolic conservation laws}.
\newblock \bibinfo{journal}{Journal of computational physics}
  \bibinfo{volume}{27}, \bibinfo{pages}{1--31}.
%Type = Misc
\bibitem[{Sousa and Scalo(2021)}]{Sousa_ArXiv_2021}
\bibinfo{author}{Sousa, V.C.B.}, \bibinfo{author}{Scalo, C.},
  \bibinfo{year}{2021}.
\newblock \bibinfo{title}{{A unified Quasi-Spectral Viscosity (QSV) approach to
  shock capturing and large-eddy simulation}}.
\newblock \bibinfo{howpublished}{(in press at the Journal of Computational
  Physics)}.
\newblock \href{http://arxiv.org/abs/2111.01929}{\tt arXiv:2111.01929}.
%Type = Article
\bibitem[{Tadmor(1989)}]{Tadmor_SIAM_1989}
\bibinfo{author}{Tadmor, E.}, \bibinfo{year}{1989}.
\newblock \bibinfo{title}{Convergence of spectral methods for nonlinear
  conservation laws}.
\newblock \bibinfo{journal}{SIAM Journal on Numerical Analysis}
  \bibinfo{volume}{26}, \bibinfo{pages}{30--44}.
%Type = Article
\bibitem[{Tadmor(1990)}]{Tadmor_NASA_1990}
\bibinfo{author}{Tadmor, E.}, \bibinfo{year}{1990}.
\newblock \bibinfo{title}{Shock capturing by the spectral viscosity method}.
\newblock \bibinfo{journal}{Computer Methods in Applied Mechanics and
  Engineering} \bibinfo{volume}{80}, \bibinfo{pages}{197--208}.
%Type = Article
\bibitem[{Tonicello et~al.(2020)Tonicello, Lodato and
  Vervisch}]{Tonicello_CandF_2020}
\bibinfo{author}{Tonicello, N.}, \bibinfo{author}{Lodato, G.},
  \bibinfo{author}{Vervisch, L.}, \bibinfo{year}{2020}.
\newblock \bibinfo{title}{Entropy preserving low dissipative shock capturing
  with wave-characteristic based sensor for high-order methods}.
\newblock \bibinfo{journal}{Computers \& Fluids} \bibinfo{volume}{197},
  \bibinfo{pages}{104357}.
%Type = Article
\bibitem[{Toro et~al.(1994)Toro, Spruce and Speares}]{Toro_SW_1994}
\bibinfo{author}{Toro, E.F.}, \bibinfo{author}{Spruce, M.},
  \bibinfo{author}{Speares, W.}, \bibinfo{year}{1994}.
\newblock \bibinfo{title}{{Restoration of the contact surface in the
  HLL-Riemann solver}}.
\newblock \bibinfo{journal}{Shock waves} \bibinfo{volume}{4},
  \bibinfo{pages}{25--34}.
%Type = Article
\bibitem[{Vreman et~al.(1995)Vreman, Geurts and Kuerten}]{vreman1995priori}
\bibinfo{author}{Vreman, B.}, \bibinfo{author}{Geurts, B.},
  \bibinfo{author}{Kuerten, H.}, \bibinfo{year}{1995}.
\newblock \bibinfo{title}{A priori tests of {L}arge-{E}ddy {S}imulation of the
  compressible plane mixing layer}.
\newblock \bibinfo{journal}{Journal of engineering mathematics}
  \bibinfo{volume}{29}, \bibinfo{pages}{299--327}.
%Type = Article
\bibitem[{Wang(2002)}]{Wang_2002_JCP}
\bibinfo{author}{Wang, Z.J.}, \bibinfo{year}{2002}.
\newblock \bibinfo{title}{Spectral (finite) volume method for conservation laws
  on unstructured grids. basic formulation: Basic formulation}.
\newblock \bibinfo{journal}{Journal of computational physics}
  \bibinfo{volume}{178}, \bibinfo{pages}{210--251}.
%Type = Article
\bibitem[{Wang and Gao(2009)}]{Wang_JCP_2009}
\bibinfo{author}{Wang, Z.J.}, \bibinfo{author}{Gao, H.}, \bibinfo{year}{2009}.
\newblock \bibinfo{title}{A unifying lifting collocation penalty formulation
  including the discontinuous {G}alerkin, spectral volume/difference methods
  for conservation laws on mixed grids}.
\newblock \bibinfo{journal}{Journal of Computational Physics}
  \bibinfo{volume}{228}, \bibinfo{pages}{8161--8186}.
%Type = Article
\bibitem[{Williams et~al.(2013)Williams, Castonguay, Vincent and
  Jameson}]{WilliamsJameson_JCP_2013}
\bibinfo{author}{Williams, D.M.}, \bibinfo{author}{Castonguay, P.},
  \bibinfo{author}{Vincent, P.E.}, \bibinfo{author}{Jameson, A.},
  \bibinfo{year}{2013}.
\newblock \bibinfo{title}{Energy stable flux reconstruction schemes for
  advection--diffusion problems on triangles}.
\newblock \bibinfo{journal}{Journal of Computational Physics}
  \bibinfo{volume}{250}, \bibinfo{pages}{53--76}.
%Type = Article
\bibitem[{Woodward and Colella(1984)}]{Woodward_JCP_1984}
\bibinfo{author}{Woodward, P.}, \bibinfo{author}{Colella, P.},
  \bibinfo{year}{1984}.
\newblock \bibinfo{title}{The numerical simulation of two-dimensional fluid
  flow with strong shocks}.
\newblock \bibinfo{journal}{Journal of computational physics}
  \bibinfo{volume}{54}, \bibinfo{pages}{115--173}.
%Type = Article
\bibitem[{Zhang(2017)}]{Zhang_JCP_2017}
\bibinfo{author}{Zhang, X.}, \bibinfo{year}{2017}.
\newblock \bibinfo{title}{{On positivity-preserving high order discontinuous
  Galerkin schemes for compressible Navier--Stokes equations}}.
\newblock \bibinfo{journal}{Journal of Computational Physics}
  \bibinfo{volume}{328}, \bibinfo{pages}{301--343}.
%Type = Article
\bibitem[{Zhu et~al.(2008)Zhu, Qiu, Shu and Dumbser}]{Zhu_JCP_2008}
\bibinfo{author}{Zhu, J.}, \bibinfo{author}{Qiu, J.}, \bibinfo{author}{Shu,
  C.W.}, \bibinfo{author}{Dumbser, M.}, \bibinfo{year}{2008}.
\newblock \bibinfo{title}{{Runge--Kutta discontinuous Galerkin method using
  WENO limiters II: unstructured meshes}}.
\newblock \bibinfo{journal}{Journal of Computational Physics}
  \bibinfo{volume}{227}, \bibinfo{pages}{4330--4353}.

\end{thebibliography}

\end{document}